\documentclass[12pt,preprint]{aastex}

\slugcomment{Accepted for publication in the Astronomical Journal}

\shorttitle{Origin \& Evol State of Hi Lat B Stars}
\shortauthors{Martin}

\begin{document}


\title{The Origins and Evolutionary Status of B Stars Found Far From
  the Galactic Plane II:  Kinematics and Full Sample
  Analysis\footnote{Based on observations made at the 2.1-m Otto
    Struve Telescope of McDonald Observatory operated by the
    University of Texas at Austin}} 


\author{J. C. Martin\altaffilmark{2}}
\affil{Case Western Reserve University Astronomy
  Department\\Cleveland, OH 44106} 
\email{jmartin@astro.umn.edu}

\altaffiltext{2}{Currently at University of Minnesota, School of
  Physics and Astronomy, Minneapolis, MN 55455} 



\begin{abstract}
This paper continues the analysis of faint high latitude B stars from
\citet{paperi}.  Here we analyze the kinematics of the stars and
combine them with the abundance information from the first paper to
classify each one.  The sample contains 31 Population I runaways,
fifteen old evolved stars (including five BHB stars, three post-HB
stars, a pulsating helium dwarf, and six stars of ambiguous
classification), one F-dwarf, and
two stars which do not easily fit in one of the other categories.  No
star in the sample unambiguously shows the characteristics of a young
massive star formed in situ in the halo.  The two unclassified
stars are probably extreme Population I runaways.  The low binary
frequency and rotational velocity distribution of the Population I
runaways imply that most were ejected from dense star clusters by DES
(dynamic ejection scenario).  However we remain puzzled by the lack of
runaway Be stars.  We also confirm that PB~166 and HIP~41979 are both
nearby solar-metallicity BHB stars.
\end{abstract}


\keywords{stars: abundances, stars:early-type, stars:evolution,
  stars:kinematics, stars: rotation, stars: Be}


\section{Introduction}
One in three faint high galactic latitude B stars have spectral
characteristics which are indistinguishable from nearby Population I B
stars \citep{gs74}.  Assuming that these are main sequence stars, their
faint apparent magnitudes imply that they are several kiloparsecs
above the galactic plane.  These stars most likely fall into one of three
categories: 
\begin{enumerate}
\item{``Normal'' massive Population I stars ejected from the galactic disk, }
\item{Misidentified older evolved stars (i.e. blue horizontal branch
  (BHB), post-HB, AGB-manqu\'e, or post asymptotic giant branch (PAGB)
  stars), or}  
\item{Young massive stars formed in situ in the galactic halo. }
\end{enumerate}

Paper~I \citep{paperi} analyzed the compositions of forty apparently
normal main sequence B stars, that are mostly 0.5 to 2 kiloparsecs from
the galactic plane, to determine to which of those three categories each
belonged.  The abundances revealed:  twenty two with Population I
abundances, nine with post-main sequence evolution, eight which could
only be characterized as metal poor, and one that was unclassifiable.
Several of those stars which did not appear to have normal disk-like
abundances also had other distinguishing characteristics including
variability and/or unique emission features.  Combining that data with
the abundances, we arrived at the following preliminary
classifications:  
\begin{itemize}
\item Runaway Population I stars:  21 stars are indistinguishable
  from nearby Population I B stars, implying that they had been
  ejected from the disk. 
\item Old evolved stars:  10 stars have abundance patterns and 
  characteristics which are expected in BHB, post-HB, AGB-manqu\'e,
  PAGB, or PPN stars  (e.g. enhancement or depletion of specific
  elements by nuclear processing, atomic diffusion, and/or depletion
  of refractory elements by dust grain formation).
\item F dwarf:  AG +03 2773 appears to have been misclassified and is
  probably an early type F dwarf 
\item Unknown:  17 stars have insufficient or conflicting information
  and cannot be categorized as runaway Population I stars or old
  evolved stars.  This category contains any stars that may have
  formed in situ in the halo.  
\end{itemize}

This paper analyzes the kinematics of each star and
combines those results with Paper~I to reclassify each star as a
Population I runaway, an old evolved star, or a young massive star
formed in situ in the halo.  Refer to Paper~I for a detailed
description of the stellar samples and the spectra.  In Section
\ref{sec2} we review the expected kinematics for each of these stellar
populations along with numerical simulations that predict the
kinematic distribution of runaway stars.  In Section \ref{sec3} we
discuss the raw data the methods used to calculate the space
velocities and in Section \ref{velocities} we analyze those
space velocities.  In Section \ref{classify} we combine the kinematic
data with the results from Paper~I to make final classifications of
each star and discuss the overall breakdown of the sample in Section
\ref{breaksec}.  Finally, in Sections \ref{discuss1} and
\ref{discuss2} we analyze the properties of the stars within each
category to infer more about the origins and characteristics of each
stellar population. 
 
\section{Background \& Theory \label{sec2}}

\subsection{An Overview of the Milky Way}
The velocities of the stars in three dimensions are calculated in the
galactocentric cylindrical coordinate system (V$_{\pi}$, V$_{\theta}$,
V$_{z}$).  In this frame of reference, V$_{\pi}$ is radial motion
relative to the galactic center with positive V$_{\pi}$ directed
away from the center of the Milky Way.  V$_{\theta}$ is the rotational
velocity component, orthogonal to V$_{\pi}$ in the galactic plane and
positive in the direction of galactic rotation.  V$_{z}$ is velocity
perpendicular to the galactic plane (parallel to the spin axis of the
galaxy), with positive V$_{z}$ towards the North Galactic Pole
(NGP).  In this system, the Sun is 8 kpc from galactic center and
moving with a velocity of (V$_{\pi}$, V$_{\theta}$, V$_{z}$) = (-9 km
s$^{-1}$, 232 km s$^{-1}$, 7 km s$^{-1}$), (the solar motion is the
sum of the motion of the Local Standard of Rest (LSR, V$_{\theta}$ =
220 km s$^{-1}$) and the Sun's own peculiar motion
\citep{mihalas81}). 

The Milky Way has three main kinematic components:  the bulge/bar, the
disk, and the halo.  None of the stars in this sample should be
associated with the bulge or bar since the influence of that
population is confined to within several kiloparsecs of the galactic
center.  Any of the stars in the sample could be members of the halo
but only those within about a kiloparsec from the galactic plane could be
considered part of the thick disk.  Table \ref{tab61} shows the generally
accepted velocity distributions of the thin disk, thick disk, and halo
populations.  The disk populations are characterized by a strong
rotational component (V$_{\theta}$) with relatively small velocity
dispersions.  The halo population, by contrast, has much larger
isotropic velocity dispersions, with no strong signature of rotation.
As a result, halo stars tend to have significantly larger V$_{\pi}$
and V$_{z}$ than most disk stars. 

The three categories of faint high latitude B stars belong to separate
galactic populations, each with their own kinematic properties and
signature.  Massive stars that formed in situ in the halo should have
halo-like kinematics which are similar to the high-velocity clouds
they formed in.  Low mass old evolved stars can belong to either the
halo or thick disk populations.  Runaway Population I stars occupy the
same volume as the halo but retain a strong rotational component to
their velocity.  This makes them a kinematically unique population
that combines the characteristics of both the disk and the halo.

\subsection{Runaway Population I Stars\label{runaways}}

There are two mechanisms that have been proposed to launch
Population~I B stars from the galactic disk:  the binary supernova
scenario (BSS; also called the Blaauw Hypothesis) and the dynamic
ejection scenario (DES).  The binary supernova scenario (BSS) was
first proposed by \citet{zwicky57} and later formalized by
\citet{blaauw61}.  In this scenario the secondary star in a binary is
ejected when the primary goes supernova and the system loses enough
mass to become unbound.  Stars launched by BSS are ejected with space
velocities comparable to their original orbital velocity (up to 200 km
s$^{-1}$; \citet{zwart00}).  Monte Carlo simulations by
\citet{zwart00} predict that in 20\% -- 40\% of BSS ejections the
neutron star produced in the supernova will remain bound to the
runaway with an orbital period of several hundred days.

\citet{poveda67} describe an alternate mechanism where stars receive a
velocity kick from close dynamic interactions between pairs of
binaries or single stars and binaries.  This scenario, known as the
dynamic ejection scenario (DES), is probably most efficient in densely
populated star clusters or star forming regions.  Simulations
by \citet{leonard90,leonard91,iben97} show that most DES encounters
should produce one or more single runaway stars with velocities up to
300 km s$^{-1}$.   

\subsubsection{Probing the Galactic Potential With Runaways}
As demonstrated in Appendix \ref{ApB}, the makeup of the galactic
potential influences the trajectories and flight times of the stars
in the sample.  Theoretically, the problem can be reversed to
constrain the galactic potential using the fact that no runaway star
has a flight time which exceeds its main sequence lifetime.  However,
in practice the uncertainty in the main sequence lifetimes is too
large to constrain the potential in any meaningful way.  We are also
limited by the degree to which components of the potential influence
individual stars.  The flight times of these stars are
mostly insensitive to the bulge component or changes in the scale
length of the disk.  Then, only stars within about 1.5 kpc of the
plane have any significant sensitivity to changes in the disk mass
surface density.  Therefore, small number statistics, lack of
sensitivity to changes in the potential, and uncertainty in the main
sequence lifetimes conspire to make runaway B stars unsuitable as
tracers for placing meaningful constraints on the galactic potential.

\subsubsection{Runaway Sample Bias\label{runawaybias}}
\begin{figure}
\figurenum{1}
\label{sampleselect}
\includegraphics[angle=90,scale=0.5]{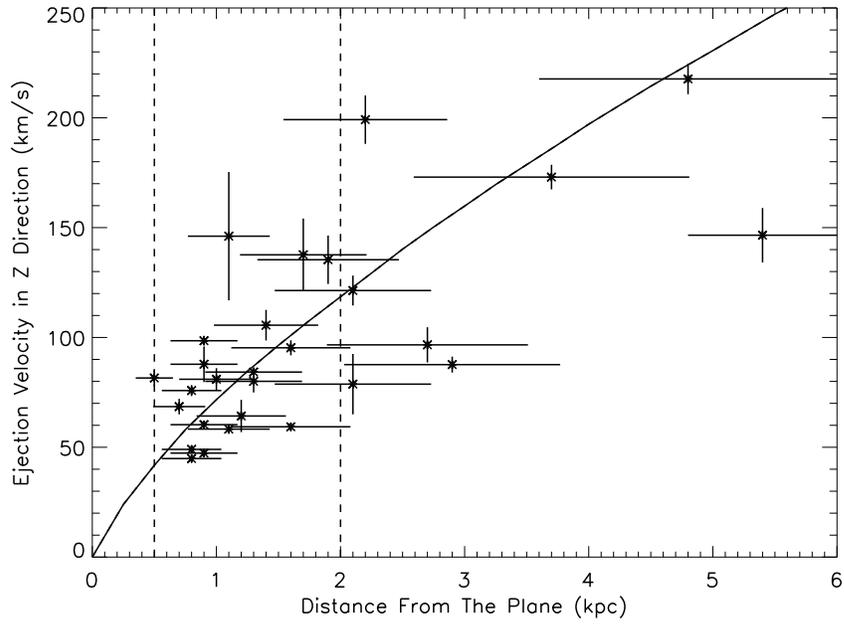}
\caption{A plot of the Population I runaway stars in the sample.  The
  solid curve is the theoretical maximum Z-distance which a star will
  reach with a given Z$_{ej}$.  The dashed vertical lines mark 0.5 kpc
  $\le$ Z $\le$ 2.0 kpc.} 
\end{figure}

If we assume main sequence luminosities and distances for all the
stars in the sample, then over 70\% of them are between 0.5~kpc 
and 2~kpc from the galactic plane (Table \ref{resultstab}, Figure
\ref{sampleselect}).  This de facto bias translates directly into a
kinematic selection effect which prevents us from using the runaways
to probe the range of ejection velocities imparted by BSS or DES.
However, it is useful to us because it gives the runaways a distinct
kinematic signature in their V$_{\pi}$ and V$_{\theta}$ space
velocities (see Section \ref{runawaysig}).

Runaways launched from the galactic disk follow a ballistic trajectory.
As a result, they spend more time near the apex of their trajectory than
traveling to or from that point.   Therefore, most runaways will be
observed near to their maximum displacement from the galactic plane.
This is why most of the runaways in our sample have relatively small
V$_z$ velocities (Table \ref{resultstab}) and are found near the
theoretical apex of their orbits (Figure \ref{sampleselect}).  The
maximum displacement from the plane is directly proportional to the
kick which a runaway receives in the V$_z$ direction (Z$_{ej}$) so
that a volume limited sample will contain mostly stars with a given
range of Z$_{ej}$.  The parameters of this sample make it heavily
biased in favor of stars with Z$_{ej}$ between 50 km s$^{-1}$ and 125
km s$^{-1}$ (assuming that the galactic potential matches the one in
Appendix \ref{galmod}).  

The relationship between Z$_{ej}$ and the total ejection velocity
(V$_{ej}$) is more complex.  We computed the sample's bias in terms of
V$_{ej}$ by Monte Carlo simulation using the galactic potential model
in Appendix \ref{galmod} and the following assumptions that model the
parameters of our sample: 
\begin{itemize}
\item The stars fall between 0.5 and 2 kpc from the galactic plane
\item Runaways are observed near the apex of their trajectory, and
\item There is no preferred direction for the orientation of V$_{ej}$
  vector in  space (all angles are equally likely).  In other words,
  Z$_{ej}$/V$_{ej}$ is equally likely to have any value between zero
  and one. 
\end{itemize}
The results of the simulation are not sensitive to the upper
bound of the first assumption.  Stars reaching a maximum
height of 3 kpc or 4 kpc have Z$_{ej}$s which are 41 km s$^{-1}$ and
79 km s$^{-1}$ faster than those reaching 2 kpc (Figure
\ref{sampleselect}).  This has only a slight affect on the high
velocity tail of the distribution.  The second assumption is merely,
as explained above, an application of Kepler's Second Law.  Figure
\ref{sampleselect} confirms that the runaways in the sample are found
near the apex of their trajectories.

\begin{figure}
\figurenum{2}
\label{selectionsimfig}
\includegraphics[angle=90,scale=0.5]{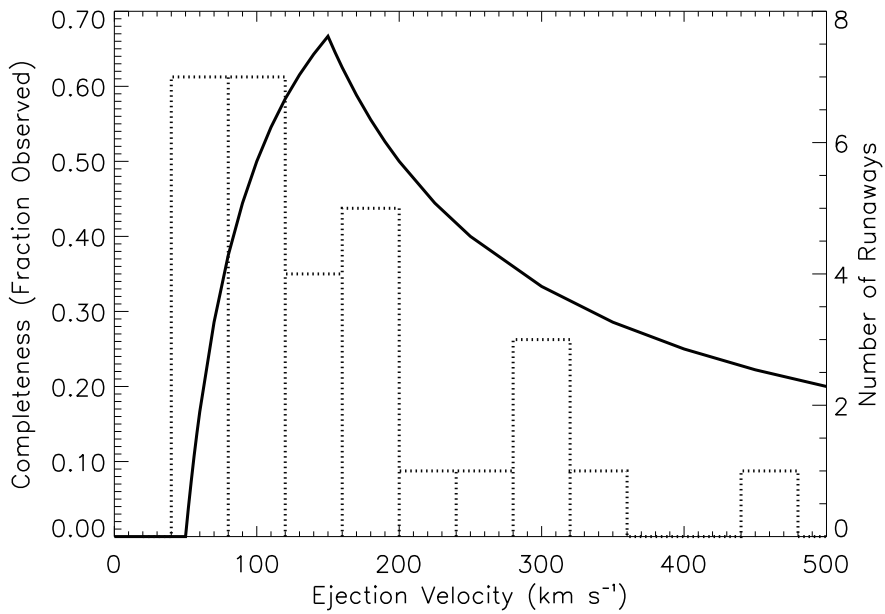}
\caption{The results of a Monte Carlo simulation of the selection
  effect with regard to ejection velocity (V$_{ej}$) in the sample
  (solid curve read off the left axis) and a histogram of the runaway
  ejection velocities in 40 km s$^{-1}$ bins (dotted line read off the
  right axis).} 
\end{figure}

The computed bias in terms of total ejection velocity is presented in
Figure \ref{selectionsimfig} along with the distribution of ejection
velocities for the runaways in our sample (see Section \ref{zej}).
The sample bias is the dominate factor shaping the distribution of
ejection velocities.  A sharp cutoff occurs between 50 km s$^{-1}$ and
100 km s$^{-1}$ because stars ejected with V$_{ej}$ smaller than 50
km s$^{-1}$ are unable to reach a height of 0.5 kpc from the plane
even if all their velocity is directed in the V$_z$ direction.  Given
this overwhelming bias, we must conclude that this sample can not
provide any useful information with respect to the distribution of
Z$_{ej}$ or V$_{ej}$.  Future studies which intend to measure the
distribution of ejection velocities must sample many more runaway
stars over a wider range of distances (i.e. 0.1 to 5 kpc) from the
galactic plane.

\subsubsection{Expected Runaway Velocities in the V$_{\pi}$ versus
  V$_{\theta}$ Plane\label{runawaysig}} 

While the sample bias is crippling with respect to the distribution of
Z$_{ej}$ and V$_{ej}$, it can be used to identify the runaways in the
sample by their space velocities.  The remainder of V$_{ej}$ that
is not in Z$_{ej}$ is distributed between the V$_{\pi}$ and
V$_{\theta}$ components.  Because the runaways are ejected from the
thin disk population, their V$_{\pi}$ and V$_{\theta}$ velocities will
be the sum of their ejection velocities and the rotation of the rest frame
in the galactic disk that they were launched from.  As a result, a
plot of V$_{\pi}$ versus V$_{\theta}$ for runaway stars should be
centered on the LSR (V$_{\pi}$= 0 km s$^{-1}$; V$_{\theta}$= 220 km
s$^{-1}$) and kinematically heated to a degree directly  influenced by
the distribution of ejection velocities present in the sample.  This
distribution contrasts with halo stars, which should have no strong
V$_{\theta}$ component and larger V$_{\pi}$ components on average.

We ran a Monte Carlo simulation of V$_{\pi}$ and V$_{\theta}$
velocities of runaway stars using the galactic potential model
described in Appendix \ref{galmod} and the following parameters: 
\begin{enumerate}
\item The runaways in the sample originate in the galactic thin disk
  within 1.0 kpc of the Sun. (see below)
\item The maximum total ejection velocity of a runaway is 400 km
  s$^{-1}$. (see \citet{iben97} and \citet{zwart00})
\item The portion of the total ejection velocity for a runaway which
  is not in Z$_{ej}$ is randomly distributed between the V$_{\pi}$ and
  V$_{\theta}$ components, with the initial pre-ejection velocity
  equal to the LSR (V$_{\pi}$= 0 km s$^{-1}$; V$_{\theta}$= 220 km s$^{-1}$). 
\item Runaways are observed at the apex of their
  trajectories. (Kepler's Second Law)
\item Runaways included in the simulation have trajectories with
  maximum heights of 0.5 kpc to 2 kpc from the galactic plane. (see
  Section \ref{runawaybias})
\item The position of a runaway at its maximum displacement from the
  galactic plane is at least 30 degrees above the galactic plane as
  viewed from the Sun. (see the sample selection criteria in Paper~I)
\end{enumerate}

\begin{figure}
\figurenum{3}
\label{vpvtsim}
\includegraphics[angle=90,scale=0.5]{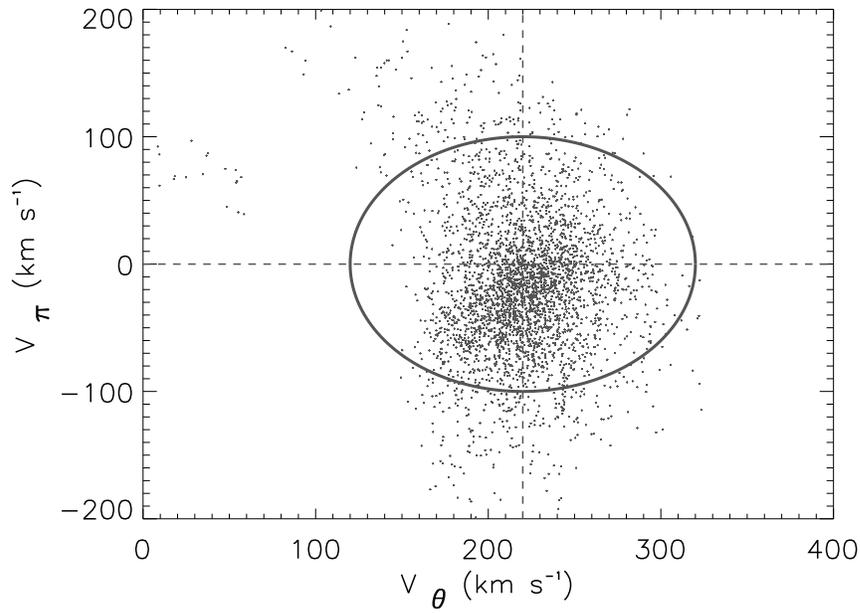}
\caption{The results of a Monte Carlo simulation of the expected
  V$_{\pi}$ versus V$_{\theta}$ distribution for stars ejected from
  the solar neighborhood included in our sample.  The dashed lines
  mark the LSR at (V$_{\theta}$,V$_{\pi}$) = (220 km s$^{-1}$, 0 km
  s$^{-1}$).  The circle centered on the LSR has a radius of 100 km
  s$^{-1}$.}
\end{figure}

The results of the simulation are presented in Figure \ref{vpvtsim}.  
Note that nearly all the runaway stars fall within 100 km s$^{-1}$ of
the LSR.  The fuzzy diagonal structure to the lower left of the LSR can be
explained by the behavior of a particle in a Keplarian potential.  An
orbiting object which receives a kick slowing its orbital velocity
(decreased V$_{\theta}$) will move radially inward (negative
V$_{\pi}$) while an object receiving a kick increasing its orbital
velocity (increased V$_{\theta}$) will move radially outward (positive
V$_{\pi}$).  This creates a noticeable diagonal pattern in the
velocity field.  While the galactic potential is not Keplarian, it is
reasonably approximated as such for the purpose of this argument. 

The second assumption further negates the significance of the upper
bound in the fifth assumption.  If more of a star's V$_{ej}$ is
partitioned into Z$_{ej}$ (to make it go further from the plane) then
there is less velocity to distribute into the V$_{\theta}$ and
V$_{\pi}$ velocity components.  This makes a {\em tighter} and more
centrally concentrated velocity distribution about the LSR in the
V$_{\theta}$ versus V$_{\pi}$ plane.

\begin{figure}
\figurenum{4}
\label{regionsims}
\plotone{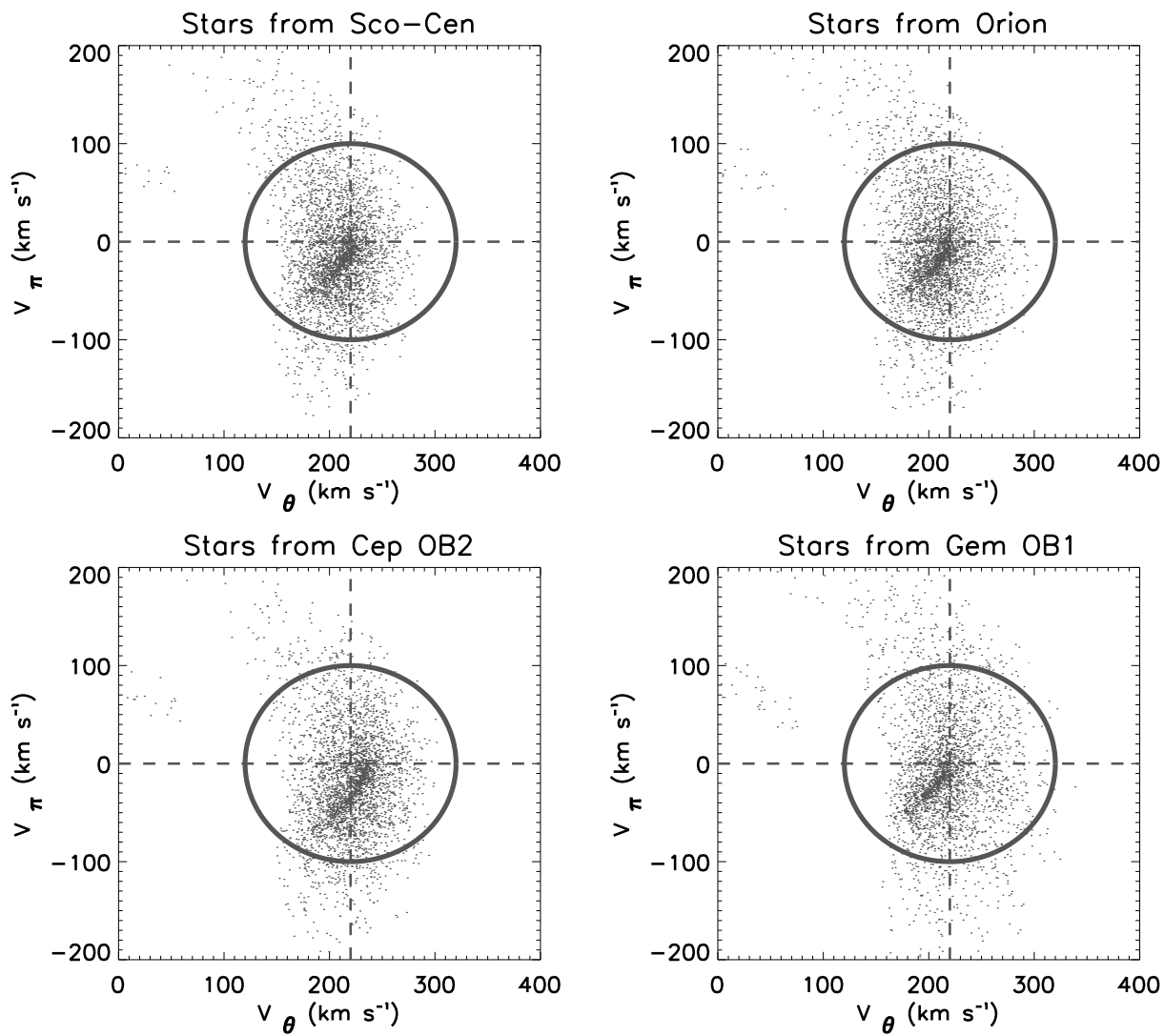}
\caption{The results of a Monte Carlo simulation of the expected
  V$_{\pi}$ versus V$_{\theta}$ distribution for stars ejected from
  specific star forming regions in the solar neighborhood.  The circle
  centered on the LSR has a radius of 100 km s$^{-1}$.} 
\end{figure}

The first assumption in the simulation also has little effect on the
outcome.  If we assume that the runaways have originated in or near
discrete local star forming regions instead, the general size and
shape of the velocity distribution remains the same (Figure
\ref{regionsims}).  There are some subtle differences between the
distributions which are dependent on the distance and location of each
star forming region relative to the Sun.  However, these differences
are small in relation to the typical observational errors associated
with the stellar velocities measured in this study and it took
thousands of points in the Monte Carlo simulation to make the
differences between these patterns noticeable.

\section{The Data \& Calculations\label{sec3}}

\subsection{Full Space Velocities}

The full space velocities of the stars are computed using the
algorithm of \citet{eggen61}.  The input parameters to this algorithm
include:  distance, proper motion, and radial velocity.  The errors in
the velocities were estimated from the errors of the input parameters
using a Monte Carlo technique.  The largest contribution to the errors
is from the proper motions because they are multiplied by the distance
and usually have errors that are a significant fraction of their
value.  Uncertainty in the other factors also contributes to the
over-all errors but to a lesser degree.

For diagnostic purposes our first assessment of each star's velocity
is made assuming that it is a main sequence star.  In doing so, we
purposefully over estimated the distances and space velocities of the
less intrinsically luminous evolved stars in the sample.  That fact is
used to distinguish them from the Population I runaways and in
Section \ref{bhbvel} we will revisit the velocities of the less
massive and luminous stars using more appropriate assumptions. 

\subsection{Proper Motions}

Every star in the study has proper motions from the Hipparcos Catalog
(HIP) \citep{hip}, the Tycho-2 Catalog \citep{tycho2}, and at least
one other catalog including:  the ACT \citep{act}, the Carlsberg
Meridian Catalogs (CMC) \citep{cmc}, and the First U.S. Naval
Observatory CCD Astrograph Catalog (UCAC1) \citep{ucac}.  Three or
more independent proper motion measures for each star were combined by
weighted average as described in \citet{martin98}.  In most cases, the
errors of the average proper motions are a factor of 1.5 to 2.0
smaller than the errors in the Hipparcos catalog (Table \ref{inputtab}). 

\subsection{Main Sequence Distances\label{sublum}}

As noted above, the distances to the stars in the sample of study are
first determined by photometric parallax assuming that all the stars
have main sequence luminosities.  
Blue horizontal branch (BHB) stars are significantly less luminous
than main sequence B stars with the same effective temperature.
Therefore the distance to a BHB star will be overestimated if the
intrinsic luminosity of a main sequence star is used to calculate its
photometric parallax.  An overestimated distance will in turn inflate
the space velocity for a star so that a sub-luminous star which is
a member of the thick or thin disk population may kinematically appear
to be a member of the halo.  In the most extreme cases, an
overestimated distance can even result in velocities that appear to
exceed the galaxy's escape velocity in the Solar Neighborhood (at
least 500 km s$^{-1}$; \citet{carney88}). 

Due to their sub-luminous nature, BHB stars in the sample are not
several kiloparsecs above the plane, but actually only a few hundred
parsecs away.  Assigning these stars main sequence distances will help
to kinematically separate them from the rest of the sample.  We will
revisit them in Section \ref{bhbvel} to evaluate their ``true'' space
velocities with more reasonable assumptions about their intrinsic
luminosities and distances.

Absolute Johnson V magnitudes were calculated for each star from the
bolometric magnitudes and bolometric corrections.  The bolometric
magnitudes were determined by matching the effective temperature,
surface gravity, and metallicity of each star the models of
\citet{schaller92}.  The bolometric corrections were interpolated
using the same parameters from a grid of models computed with
ATLAS12\footnote{
  Grids computed by ATLAS12 using up to date opacities and atomic data
  are available online from {\tt http://kurucz.harvard.edu/grids}.}
\citep{kurucz}. 

We compared the absolute magnitudes from this method with those
calculated by the method of \citet{cramer99} for the stars in our
sample with Geneva Photometry (see Paper~I).   The \citet{cramer99}
relation was calibrated using Hipparcos parallaxes to obtain the
distances and intrinsic luminosities of stars fitted as a function of
the Geneva reddening free X and Y color indices \citep{cramer79}.
There are fewer stars with Hipparcos parallaxes to calibrate the hot
end of the relation (T$_{eff}$ $\ge$ 30000~K) so one must be cautious
applying this relation in that regime.  This method is also only valid
for stars with effective temperatures hotter than 11000~K because it
uses the X and Y indices.  

\begin{figure}
\figurenum{5}
\label{genevay}
\includegraphics[angle=90,scale=0.5]{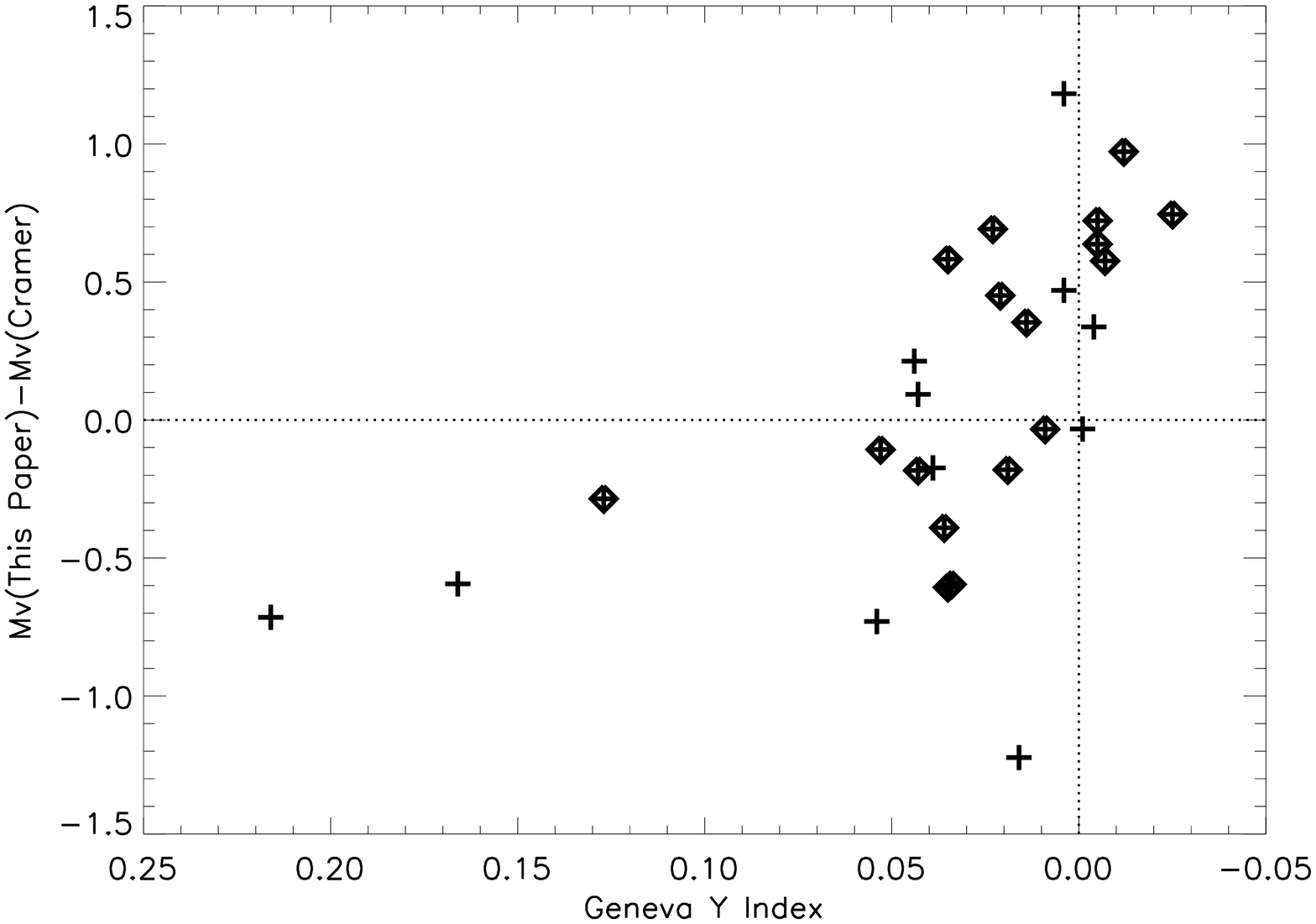}
\caption{The difference between the absolute magnitudes derived by our
  method and \citet{cramer99} versus the Geneva Y index for each
  star.  Stars which are later classified as Population I runaways are
  highlighted with diamond shapes.}
\end{figure}

There is a trend in the difference between the absolute
magnitudes calculated by our method and the method of Cramer with
respect to the Geneva Y index (Figure \ref{genevay}).  The outliers from
the main group in that diagram are stars which are obviously not main
sequence stars that lie outside the normal regime of the Cramer
relation.  The Geneva Y index is a measure of a B star's separation
from the main sequence \citep{cramer79}.  Therefore the trend in Y
suggests that the Cramer relation does not completely account for
changes in brightness as a star evolves away from the zero-age main
sequence (ZAMS).  The Cramer relation was developed from the limited
number of B stars with measurable Hipparcos parallaxes so it is
afflicted by small number statistics and strongly biases by galactic
structures such as Gould's Belt\footnote{Gould's Belt
  \citep{gouldsbelt} is a structure in the   solar neighborhood
  roughly 1 kpc in diameter that is composed of   stars in a ring of
  OB associations that are all close to the same   evolutionary age
  \citep{gbeltage}.}.  Many of the stars in this sample are at least
several tens of Myr evolved from the ZAMS (as evidenced from their
flight times, see Section \ref{zej}) so that evolution has had a
significant effect on their absolute luminosity.

The Cramer relation has an internal formal error of 0.44 mag which
corresponds roughly to a factor of 20\% in distance.  Considering our
sources of error, those of \citet{schaller92}, and the random spread
in our comparison with the \citet{cramer99} relation,  the photometric
distances are assigned a $\pm$30\% error (0.57 magnitudes). 

HD~138503 is an eclipsing binary composed of two main sequence B stars
that was analyzed by \citet{itlib}.  From that work, a distance of
2.4~kpc is adopted with an error of $\pm30\%$.  This distance is
significantly greater than the distance derived from our method
described above ($1.3\pm0.4$ kpc). However this discrepancy is
explained by HD~138503's binary nature which makes the pair more
luminous together than a single star at the same distance.

In Paper~I, we showed that AG~+03~2773 is probably a rapidly rotating
F~dwarf.  To calculate its distance we assumed an effective
temperature of 7000~K with a surface gravity of 4.0 and roughly solar
composition.  This yielded a distance of 400 pc which is consistent
with its Hipparcos parallax (0.23$\pm$2.05).  

\subsection{Radial Velocities}

Radial velocities were measured for forty (41) of the stars in the
sample.  The high resolution (R=60000) spectra taken for the abundance
analysis were used to measure the radial velocities of the stars from
the Doppler shift of their photospheric absorption lines.  (For a
detailed description of the spectra see Section 2.1 in Paper~I.)  The
resolution of the spectrograph (one resolution unit equals two pixels)
is about 2 km s$^{-1}$ but the FFT interpolates four points to each
pixel allowing an even higher precision to be achieved.  The precision
of the radial velocities is primarily influenced by our capability to
determine the line center.  This becomes more difficult with an
increase in rotational broadening.

The process of measuring the wavelengths of line centers was
interactive.  Lines that appeared to be blended or had no clear
minimum in the profile were excluded from the analysis. The line
centers are defined by the minima in the line profiles in the
continuum corrected FFT filtered data.  High rotational broadening and
low signal to noise sometimes drastically limited the number of lines
that could be measured in a spectrum.  In a three desperate cases (HD
15910, HIP~12320, and HD~106929), where no other lines could be
measured with confidence, only the hydrogen Balmer lines were used to
determine the radial velocity of the star. Because the spectral orders
have some wavelength overlap, several lines were measured on two
adjacent spectral orders.  Each of these were treated as independent
measures when computing the average value.  The average radial
velocity for a spectral observation is calculated as a straight
unweighted average of the values from each of the observed lines.
Since the distribution of individual measurements is roughly Gaussian,
the error of that average is estimated from the standard deviation of
the measurements. 

The IRAF procedure noao.rv.rvcorrect was used to transform the
observed radial velocities into the heliocentric reference frame using
the date and time at mid-exposure.  For spectra that consisted of
combined exposure sets from several nights, the correction factor was
not significantly different for each exposure set and an average value
was used. 

Radial velocity standards from \citet{fekel99} were used to verify the
measured radial velocities.  Table \ref{stdrvtab} shows that this
method is able to accurately reproduce the radial velocities of the
standard stars.  There is good agreement between most of the
radial velocities we measured (Table \ref{inputtab})  and those
published in the literature (Table \ref{rvcomp}).   Later in Section
\ref{binaries} we will discuss if any of the discrepancies in Table
\ref{rvcomp} imply that some of these stars are binaries.  

The radial velocities of ten stars were taken from the literature for
the following reasons:
\begin{itemize}
\item{No spectra were obtained for six stars in the sample.}
\item{HD~140543 has irregularly shaped lines that proved difficult to
  measure with any confidence.}
\item{The systemic velocity of BD~+13 3224 could not be determined
  with any confidence because the lines were smeared (the exposures
  were too long compared to the very short period of this pulsating
  helium dwarf).}
\item{HD~135569 is a long period binary so we relied on the analysis
  of \citet{bolton80} for the systemic  velocity.}
\item{The systematic radial velocity of eclipsing binary HD~13503 
(IT Lib) was analyzed separately by \citet{itlib}.}
\end{itemize}

\subsection{Main Sequence Lifetimes\label{mslife}}

The projected main sequence lifetimes (T$_{MS}$) for the stars in the
sample were determined from their effective temperature, surface
gravity, and metal content from the tables of \citet{schaller92}.  If
the metallicity of the star is not know, the longest main sequence
lifetime for any metallicity is used. 
The results for each star are listed in Table \ref{resultstab}.  
The error in the main sequence lifetime is 
difficult to estimate since stellar rotation and other mixing can
affect the hydrogen burning lifetime of a massive star by as much as
30\% \citep{heger00}.  Other processes (i.e. convection,
overshooting, and semiconvection) are only roughly modeled and
magnetic fields are completely ignored, affording significant
additional uncertainty \citep{hegeretal00}.  Considering these
factors and the uncertainty of converting the effective temperatures
and metallicities to main sequence lifetimes, we have adopted an error
of $\pm$50\% for the estimated main sequence lifetimes.  The error
could be larger than this for the highest mass stars (shortest
main sequence lifetimes).

\subsection{Ejection Velocities and Flight Times\label{zej}}

Initial ejection velocities perpendicular to the plane (Z$_{ej}$) and
travel times back to the disk (T$_{flight}$) are computed from the
full space velocities of each star (assuming main sequence distances)
using a model of the galactic potential and an orbital integrator
outlined in Appendix \ref{galmod} and described in detail by
\citet{harding01}.  Errors in Z$_{ej}$ and T$_{flight}$ were estimated
from the errors in each star's position and space velocity by
Monte-Carlo simulation.   

The time of flight for each star is measured from the star's present
position to Z = 0 along its orbit in the galactic potential.  Stars
that are moving back toward the galactic plane are traced back through
the apex of their orbit.  It is reasonable to assume that each star
began at Z=0 since most star forming regions are found very close to
the galactic plane.  However, a few like the Orion complex are up to
100 pc from the galactic plane.  In those cases the difference between
the actual launch point and Z = 0 has negligible affect on the flight
times because the stars are moving with velocities on the order of
hundreds of km s$^{-1}$ when they are close to the plane and traverse
100 pc in less than 1 Myr (well within the margin of error).  The
flight times and Z$_{ej}$ for the sample stars are given in Table
\ref{resultstab}.

\section{Analysis of Space Velocities\label{velocities}}

Table \ref{resultstab} lists the spatial coordinates, full space
velocities, main sequence lifetimes, ejection velocities, and flight
times for the stars in the sample.  A number of stars, PB~166,
HD~218970, BD~+33~2642, and HIP~1511 in particular, stand out as
having unusually large velocities relative to the LSR.  As noted in
Section \ref{sublum}, these extreme velocities may be the result of
overestimating the distance to the star.  The V$_z$ for twenty two of
the stars is under 50 km s$^{-1}$.  As noted in Section
\ref{runawaybias}, most runaways are observed near the apex of their
trajectory so they should have V$_z$ close to zero.  Twelve (12) of
the stars are actually moving back toward the galactic plane. If they
are runaways, then these stars have already passed through the apex of
their trajectory and are falling back toward the galactic disk. 

\subsection{V$_{\pi}$ versus V$_{\theta}$}

\begin{figure}
\figurenum{6}
\label{vpivsvtheta} 
\includegraphics[angle=90,scale=0.5]{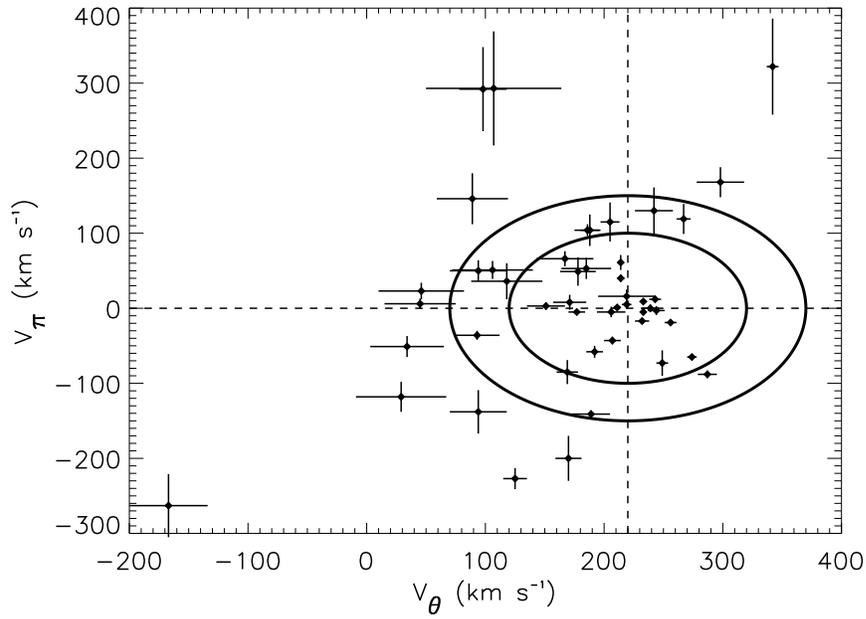}
\caption{A plot of V$_{\pi}$ versus V$_{\theta}$ for stars in the
  sample assuming main sequence distances.  The dashed lines mark the
  LSR at (V$_{\pi}$,V$_{\theta}$) = (0 km s$^{-1}$, 220 km s$^{-1}$).
  The circles centered on the LSR are draw with radii of 100 km
  s$^{-1}$ and 150 km s$^{-1}$.} 
\end{figure}

A plot of the V$_{\pi}$ velocity versus V$_{\theta}$ velocity for
stars in the sample (Figure \ref{vpivsvtheta}) shows that there is a
significant clustering about the LSR (V$_{\theta}$ = 220 km s$^{-1}$,
V$_{\pi}$ = 0 km s$^{-1}$) which is similar to the simulated runaway
star distribution (Section \ref{runawaysig}, Figure
\ref{vpvtsim}).  The region within 100 km s$^{-1}$ of the LSR contains
twenty four stars.  These are the most likely runaway candidates in the
sample.  That is not to say that all twenty four stars in that region
are definitely runaways but rather that this is one piece of evidence
in favor of that classification.  Likewise, some of the stars between
100 km s$^{-1}$ and 150 km s$^{-1}$ of the LSR are not completely
excluded from being runaways but stars more than 150 km s$^{-1}$ from
the LSR are most likely not runaways. 

\subsection{Main Sequence Lifetime versus Travel Time}

Figure \ref{msvstd1} compares the estimated main sequence lifetime
to the calculated flight time for each star in the sample (assuming
main sequence distances).  Between the dotted lines the main sequence
lifetime of a star is within $\pm50\%$ of the flight time (the
estimated error, see Section \ref{mslife}).  These lines divide the
diagram into three regions:  stars with flight times significantly
longer than their project main sequence lifetimes (upper left), stars
with flight times about equal to their main sequence lifetimes
(between the lines), and stars with flight times which are definitely
shorter than their main sequence lifetimes (lower right). 

The upper left corner of the diagram contains stars that could have
formed in situ in the halo or sub-luminous stars whose distances,
velocities, and flight times have been overestimated.  Only one star
is from our sample is found in this part of the diagram:  HD~140543.
In that case it is possible that the error in the main sequence
lifetime is greater than 50\% so that its liftime and travel time
could be consistent.  It is uncertain if there is any conflict between
the flight times and hydrogen burning life times of the fifteen stars
which fall between the dotted lines.  The remaining thirty three stars
to the right of the dotted lines could easily have traveled from the
disk to their present location in their main sequence lifetimes. 

For comparison, we have highlighted the most likely runaway stars from
Figure \ref{vpivsvtheta} (stars within 100 km s$^{-1}$ of the LSR).  All
but HD~140543 fall between or to the right of the dotted
lines, confirming that they are most likely stars which have been
ejected from the galactic disk.   

\section{Kinematic Scoring and Classification\label{classify}}

\begin{figure}
\figurenum{7}
\label{msvstd1}
\includegraphics[angle=90,scale=0.5]{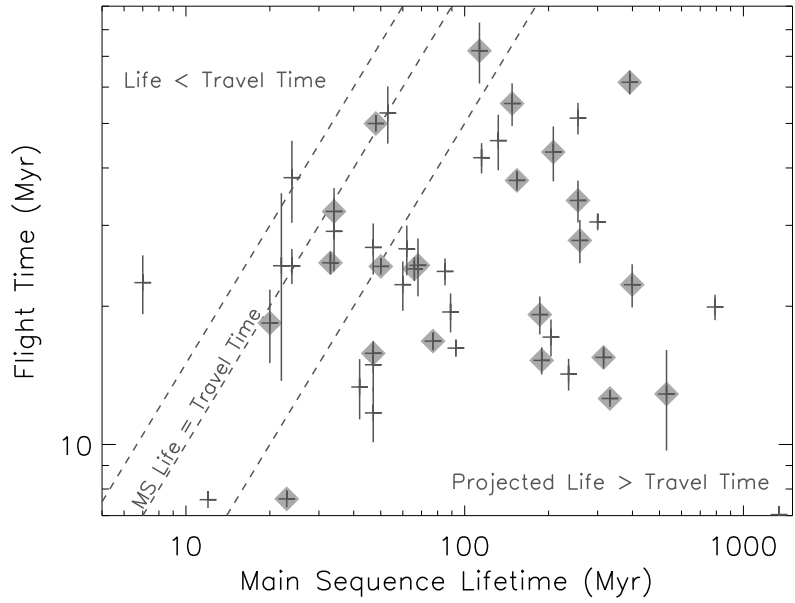}
\caption{The flight time (T$_{flight}$) from the disk (assuming main
  sequence distances) plotted against the estimate of the main
  sequence lifetime.  Stars inside the circled region from Figure
  \ref{vpivsvtheta} are highlighted with a filled diamond.} 
\end{figure}

A four point scale is used to assign a kinematic score to each
star in the sample (Table \ref{tab82}).  The points are assigned to
each star based on its velocity in the V$_{\pi}$ versus V$_{\theta}$
plane assuming main sequence distances (Figure \ref{vpivsvtheta}) and
a comparison between its flight time and main sequence lifetime
(Figure \ref{msvstd1}) as follows: 
\begin{itemize}
\item{One point is awarded to each star within 150 km s$^{-1}$ of the
  LSR in the V$_{\pi}$ versus V$_{\theta}$ plane.} 
\item{An additional point is awarded if the star is less than 100 km
  s$^{-1}$ from the LSR in the V$_{\pi}$ versus V$_{\theta}$ plane.} 
\item{One point is awarded to each star in Figure \ref{msvstd1} with a
  main sequence lifetime which is within 50\% of their flight time
  (stars between the dotted lines).} 
\item{Two points are awarded to each star with a main sequence lifetime
  at least 50\% greater than their flight time (stars to the right
  of all dotted lines).}   
\end{itemize}
In this scheme, a cumulative score of four represents a star with
kinematics that are completely consistent with a runaway from the disk,
while a score of zero denotes a star that is probably not a runaway.
A high score does not ensure that a star is a runaway but a lower
score lessens the likelihood that it was ejected from the disk.  

\subsection{Sample Breakdown and Classification}
Table \ref{tab83} combines the kinematic results from the preceding
section with the abundance and spectral flux information from Paper~I.
These data are used to classify each star as a runaway Population I
star, an old evolved star, or a massive young star formed in situ in
the halo.  While all the data are considered, the abundances are given
more weight than the other factors because we have a higher
degree of confidence in those results.  
A question mark (?) after the
classification denotes some uncertainty due to missing or conflicting
data.   

Two stars are labeled ``unknown'' (U) because they have kinematics
that are somewhat inconsistent with Population I runaways but no other
conclusive evidence as to their nature (see Section \ref{noclass}).
Also recall from Paper~I that AG +03 2773 is probably a misclassified
F Dwarf.  The remaining 46 stars can be conclusively sorted as
either old evolved stars (15 stars) or Population I runaways (31
stars).  {\em No star in this sample unambiguously shows the
  characteristics of a young massive star which formed in situ in the
  galactic halo.} 

\subsection{Breakdown of Old Evolved Stars}

\begin{figure}
\figurenum{8}
\label{oestg}
\includegraphics[angle=90,scale=0.5]{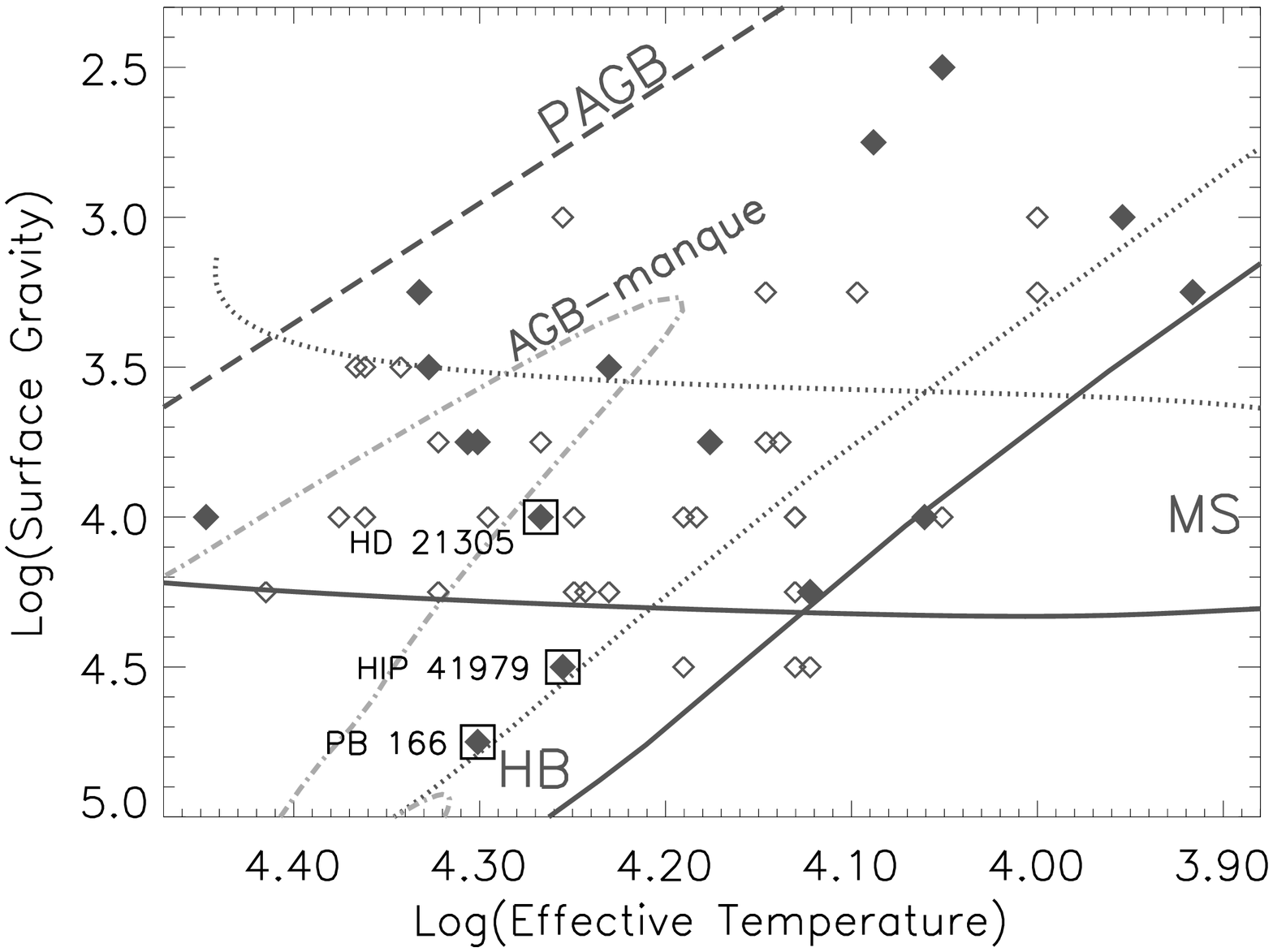}
\caption{A plot of the effective temperatures and surface gravities of
  the post-main sequence program stars (filled diamonds) and runaway
  main sequence star (smaller open diamonds) compared to the
  position of the zero age (solid line) and terminal age (dotted line)
  Main Sequence \citep{zams,tams}, the zero age (solid line) and
  terminal age (dotted line) Horizontal Branch \citep{dorman93}, a
  representative post-asymptotic branch track for a star of
  0.6 M$_{\sun}$ {\it after} AGB mass-loss (dashed line) \citep{pagbtrack},
  and a representative AGB-manqu\'e track from the \citet{dorman93} E64
  model starting with a HB mass of 0.505 M$_{\sun}$ (light dash-dot
  line).  The tracks are for solar metallicity ([Fe/H] = 0) except the
  AGB-manqu\'e track which is [Fe/H] = -1.48. The stars HD~21305, HIP
  41979, and PB 166 are highlighted for Section \ref{bhbmetal}.} 
\end{figure}

The stars classified as old evolved stars can be further sub-divided
into BHB and post-HB stars.  For the purpose of this study, a post-HB
star is any star that has evolved past the horizontal branch.  These
include PAGB, AGB-manqu\'e, and PPN (proto-planetary nebula) stars which
are collectively known as UV-bright stars in galactic globular clusters.
There are a number of differences between BHB and post-HB stars: 
\begin{itemize}
\item{The atmospheres of BHB stars are dominated by the affects of
  atomic diffusion \citep{michaud83} and polluted by the
  products of partial CNO burning while post-HB stars can have much more
  complicated abundance patterns involving more advanced nuclear
  process and dust formation since they are more evolved.} 
\item{BHB stars have low rotational velocities so that their vsin(i)
  values should be less than 40 km s$^{-1}$ \citep{peterson83} while there
  are no established observational constraints on the vsin(i) values
  of post-HB stars. }
\item{BHB stars should have effective temperatures and gravities which
  place them on or near the horizontal branch on a theoretical
  HR-diagram.  Post-HB stars can follow a number of tracks encompassing a
  wider range of gravities and effective temperatures (Figure \ref{oestg}).} 
\item{BHB stars are sub-luminous with respect to main-sequence stars
  so they will have received low kinematic scores in Table
  \ref{tab82}.  This is not necessarily so for post-HB stars.}
\end{itemize}
Given these differences we conclude that five of the old evolved stars
are BHB stars, three are post-HB stars, and six are of ``ambiguous''
classification (Table \ref{oesclass}).
Any uncertainty in the classification a star as OES in Table 6 is
propagated forward as an ``?'' appended to this classification.
BD~+13~3224 is a special case which is neither BHB or post-HB.
\citet{saio00} conclude that this helium-rich pulsator is the product
of a merger of two helium white dwarfs.  We defer to their more
exhaustive analysis.

The case of BD~+33~2642 is explored in detail in Appendix~A of
Paper~I.  This star is known as a halo PPN
\citep{napiwotzki93,napiwotzki94} with both classic nebular emission
lines and a stellar photospheric absorption spectrum.  It is iron
deficient with enhancements of carbon, nitrogen, oxygen, and
$\alpha$-process elements \citep{napiwotzki94,napiwotzki01,paperi}.
The carbon and silicon abundances are very sensitive to the surface
gravity and effective temperature used in the analysis \citep{paperi}.
Abundances of those elements are critical to deciphering the role of
mixing, nuclear processing, and refractory grain formation in the
stellar atmosphere.  Even though BD~+33~2642 is clearly a post-HB star
with the beginnings of a planetary nebula, these uncertainties leave
many open questions regarding its exact evolutionary status.

There is also notable uncertainty in the evolutionary status and
classification of the stars with vsin(i) greater than 50 km
s$^{-1}$.  These stars may be post-HB but they are classified as
``ambiguous''.  In these cases, rotational broadening hampers the
abundance analysis because only the strongest lines can be measured
and those may be blended.  HD~40267, HD~105183, and BD~+36~2268 all
have reasonably well determined CNO abundances which show their
photospheres have been significantly contaminated by fusion products.
It is possible that these stars are normal Population I objects which
were severely polluted by a nearby supernova a la BSS.  However,
BD~+36~2268 and HD~105183 also have depressed silicon abundances which
are well outside the range observed in normal Population I B stars.
The photospheric abundances for HD~1112 are a confusing jumble so the
most definitive thing we can conclude is that it is abnormal.  

HD~233622 and HD~237844 are the most rapid rotators in the OES sample.
They have silicon abundances, which like other OES stars, are
depressed well below the normal range for B stars.  Additionally
HD~233622 has an kinematic score of 1 and its oxygen abundance (from
two reliable lines) is raised slightly relative to the B star control
sample.  Unfortunately, we were not able to measure any CNO abundances
for HD~237844 but it has an Si/S ratio which is exceptional.  The
abundances of both stars and the kinematic score of HD~233622 indicate
that they are OES.  However, their high vsin(i)'s imply that they
cannot be low mass stars.  

Table \ref{oesclass} includes the critical rotational
velocities\footnote{The rotational velocity at which the centripetal
  acceleration at the equator is equal to the surface gravity of the
  star.  v$_{crit}=\sqrt{\frac{GM}{R}}$} 
calculated for the OES stars using the mass and radius from the tracks
of \citet{dorman93} (see Table \ref{bhbkin}).  Both HD~233622 and
HD~237844 have vsin(i)'s that are comparable to their critical
rotation velocity.  It is unlikely that a star so close to its
critical velocity could exist in any stable configuration. However, if
they are higher mass Population I stars (of 6 or 7 M$_\sun$, see Table
\ref{inputtab}), then their critical velocities would be much higher
and there would be no conflict.  

HD~237844 has a space velocity consistent with ejection from the disk.
However, HD~233622 has velocities which are much more extreme.  Both
have low silicon abundances which indicate that they could be low
metallicity stars formed in situ in the halo.  But their other
abundances show additional anomalies and more evidence is needed to
support such an extraordinary claim.

\subsection{The Stars We Cannot Classify\label{noclass}}

Two of the stars in our sample (HD~121968 and HD~125924) have received
a classification of ``unknown'' (U) due to insufficient information.
Both have kinematics that are not consistent with Population I
runaways (kinematic scores of 2; see Table \ref{tab82}) and no spectra
or abundances.   It is conceivable that they could be metal poor
Population I stars formed in situ in the halo or BHB stars which have
had their kinematics exaggerated (see Section \ref{sublum}).  However,
since there is no significant discrepancy between the flight times and
main sequence lifetimes of these stars it is far more likely that the
conclusions of \citet{conlon90} and \citet{conlon92} are correct and
these are runaways ejected with enough force to accelerate their
V$_{\theta}$ and V$_{\pi}$ far from the LSR. 

\subsection{Overlap With Other Studies}

A number of the stars in our sample have been categorized individually
or in small groups by other studies.  In most cases those
classifications agree with the classifications made here. The are:
BD~-15~115 \citep{ramspeck01b,magee01,conlon92}, Feige~40
\citep{conlon89a}, HD~100340 \citep{ryans99,keenan87,conlon89a},
HD~105183 \citep{dufton93}, PB~166 \citep{deboer88,conlon89a},
Feige~84 \citep{lynn04,saffer97}, HD~123884 \citep{bidelman88}, HD~137569
\citep{danziger70}, HD~138503 \citep{itlib}, BD~+33~2642 (See Paper~I,
Appendix A), HD~146813 \citep{conlon89a,conlon88}, HD~149363
\citep{mdzinarishvili05}, BD~+13~3224 \citep{jeffery01,jeffery86},
HD~206144 \citep{keenan82}, HD~218970 \citep{conlon89a,conlon89}, and
HD~220787 \citep{keenan82}.  

Our classification disagrees with those given for BD~+36~2268
by \citet{conlon89} and HD~149363 by \citet{zboril00}.  We disagree with
\citet {zboril00} over the proper stellar atmospheric parameters for
HD~149363 (for details see Paper~I).  For BD~+36~2268, we agree with
\citet{conlon89} on the parameters we use in the abundance
analysis and that there is no conflict between its flight time and
main sequence lifetime.  We also agree that the C abundance is
depressed while N \& O are ``normal.''  Taken with our Al, Si, and S
data, this indicates to us that BD~+36~2268 is a hot post-HB star rather
than a Population I B star as asserted by Conlon et al.

Several other studies overlap with a substantial fraction of our
sample.  Ten of our stars were identified by \citet{keenan83} as
Population I runaways based on their flight times.  We agree
that there are no significant conflicts between the flight times and
main sequence lifetimes for any of those stars.  However, we have
classified two of them (HD~105183 and BD~+36~2268) as old evolved
stars based on their abundances.  

Sixteen stars in our sample were also cited by \citet{conlon90} as
stars with space velocities that imply they were ejected from
the galactic disk.  We agree that thirteen of these are Population I
runaways and we have classified one other which could be a runaway as
``unknown'' (see Section \ref{noclass}).  The remaining two are
classified here as post-HB stars.  While they have flight times that are
consistent with their main sequence lifetimes, they also have peculiar
abundance patterns (HD~105183 and BD~+36~2268).

\citet{allen04} also assess the kinematics of sixteen stars in our
sample.  We only disagree with their assessment of HD~206144 which
according to them has a flight time that significantly exceeds its
main sequence lifetime.  Our analysis shows that there is no
inconsistency between the values we calculated for that star.  It is
unclear how \citet{allen04} computed their flight times or what model
for the galactic potential they used so we do not know how our analysis
differed from theirs.

\citet{behr03} classified seventeen stars in this study 
according to their [Fe/H] and
position relative to evolutionary tracks on a T$_{eff}$ versus log(g)
diagram.  The effective temperatures and surface gravities derived by
Behr closely match the values that we have adopted (Paper~I).  There
are no conflicts between our classifications and Behr's for ten of
the common stars. 

We have classified HD~1112, HD~233622, HD~105183, and BD~+36~2268 as
``ambiguous'' old evolved stars while Behr classified them as ``main
sequence.''  It is difficult to tell the difference between a main
sequence and post-HB star on a T$_{eff}$ versus log(g) diagram.  Our
classification is primarily based on the abundance ratios of elements which
\citet{behr03} did not measure.  In addition to that, each of these stars
either have only marginal agreement between their flight times and
main sequence lifetimes (BD~+36~2268) or large V$_{\pi}$
and/or V$_{\theta}$ velocities assuming main sequence distances
(HD~233622, HD~1112, and HD~105183) which imply that they are not
runaway Population I stars. 

\citet{behr03} classified BD~-7~230, BD~+30~2355 and HD~213781 as
``possible HB'' because their T$_{eff}$ and log(g) are consistent with
the horizontal branch and they have small vsin(i).  We have classified
each of these stars as Population I runaways based on their
abundances.  BD~-7~230 is slightly metal poor ([Fe/H]$\sim$-0.4),
but this is not outside the range of values observed in the control
sample of nearby Population I B stars.  It also has a normal Mg/Fe
ratio while every other BHB star in this sample is abnormal in that
respect.  In Paper~I we discussed how a difference in the
microturbulent velocity may have led \citet{behr03} to significantly
underestimate [Fe/H] for BD~+30~2355 but otherwise we are in agreement
with all the parameters that Behr derives for these stars.  The
abundances of these stars are not inconsistent with their
classification as runaways and their kinematics assuming main sequence
distances are very consistent with that population (each received a
kinematic score of 4, see Section \ref{classify}).  Whereas if these
were BHB stars there would be noticeable inconsistencies in their
space velocities from overestimating their distances (see Section
\ref{sublum}).

\section{The Breakdown of High Latitude B Stars\label{breaksec}}

Our sample is roughly magnitude limited by virtue of having been selected
from the Hipparcos catalog (which is complete up to V $\sim$ 10) and
the color-magnitude limited sample of \citet{gs74} (see Section 2,
Paper~I).  As such, the overlap between this study and \citet{gs74}
(32 stars with an FB number in Table 1 of Paper~I) represent a sample
that can be used to make rough conclusions about the proportion of
Population I runaways and older evolved stars among the faint B stars
at high galactic latitudes. 

Table \ref{breakdown} compares our sample with the results from other
similar surveys which like ours are roughly limited to stars with
V magnitudes between 9 and 12.  All these studies use abundance analysis and
kinematics to classify the stars in their samples.  \citet{rolleston97}
use radial velocities alone while \citet{magee01},
\citet{ramspeck01a}, \citet{ramspeck01b}, and our own study examined
the full space velocities.  Note that \citet{rolleston97} and
\citet{magee01} are part of a comprehensive survey, while
\citet{ramspeck01a} and \citet{ramspeck01b} are randomly selected
samples.

The only overlap between these studies and our own is runaway star
BD~-15~115 in \citet{ramspeck01b} and \citet{magee01}.  All of us find
comparable proportions of runaway Population I stars and old evolved
stars with no evidence of Population I stars formed in situ in the
halo.  There is remarkable agreement between our study
and \citet{magee01} and \citet{rolleston97}.  The general consensus
appears to be that about two thirds of all high galactic latitude B
stars are high mass Population I runaways, while the remaining third
are low mass older evolved stars.  

\section{Properties of the Population I Runaways\label{discuss1}}

\subsection{Runaway Binaries\label{binaries}}
The fraction of runaways with binary companions depends on the
ejection mechanism which accelerated them.  Simulations predict that
10\% of stars ejected by DES are binary\citep{leonard90}.  The runaway
binaries produced by DES are pairs of normal main sequence stars
observed as double lined spectroscopic binaries (SB2s).  Whereas,
binaries with captured neutron star companions should account for
20\%--40\% of runaways ejected by BSS\citep{zwart00}.  Therefore, a
significant number of BSS runaways should be single-lined
spectroscopic binaries (SB1s) with periods on the order of a few
hundred days \citep{zwart00}.  

Almost all of the stars in this sample were observed at least twice
over the course of a few days so the radial velocity measurements are
sensitive to variations in short period binaries.  The level of
sensitivity is comparable to the errors given for the radial velocity
in Table \ref{inputtab}.  They ranged from 2 km s$^{-1}$ -- 15 km
s$^{-1}$ depending on the number of lines measured and the vsin(i) of
the star.  We detected the eclipsing binary HD~138503 \citep{itlib}
but found no other short period binaries in our sample.  HD~138503 was
probably produced by DES since it is an SB2 composed of two main
sequence stars. 

In the case of five old evolved stars (HD~21305, HD~40267, HIP~41979,
HD~237844, and HD~137569) and seven runaways (HD~21532, BD~+61~996,
HD~103376, BD~+30~2355, HD~146813, HD~149363, and HD~206144) the 
observations were spread over several months so that they are
sensitive to variations from longer period variables.  We detected none
except the known binary old evolved star HD~137569. 

Finally, we compared the measured radial velocities with the
literature (Table \ref{rvcomp}).  Five runaways have radial velocities
which differ from their literature values by more than two sigma:
HD~15910, Feige~40, HD~110166, HD~188618, and HD~216135.  The radial
velocities for HD~15190, Feige~40, and HD~110166 have very large
errors so any discrepancy has a lower significance.  HD~188618 and
HD~216135 disagree rather significantly with previously published
values. They are the best candidate SB1s in our sample.  We are unable
to verify that any of these are binaries. 

The binary frequency observed by \citet{gies86} predicts about
three binaries in a sample of 28 runaways.  We found HD~138503 plus
two promising candidates.  Furthermore, those two promising candidates
are most likely long period SB1s of the type expected for stars ejected by
BSS\citep{zwart00}.  However, neither show any evidence that their
photosphere was polluted by supernova ejecta (see Paper~I). 

Our results, taken by themselves, are not significant.  However,
a number of more extensive surveys have determined that fewer than
10\% of early type high latitude runaways have binary companions
\citep{gies86,philp96,sayer96}.  That binary frequency is more
consistent with simulations of DES than BSS.

\subsection{Rotation of Runaway Stars \& the Ejection Mechanisms\label{runawayrot}}
\begin{figure}
\figurenum{9}
\label{slowrot}
\includegraphics[angle=90,scale=0.5]{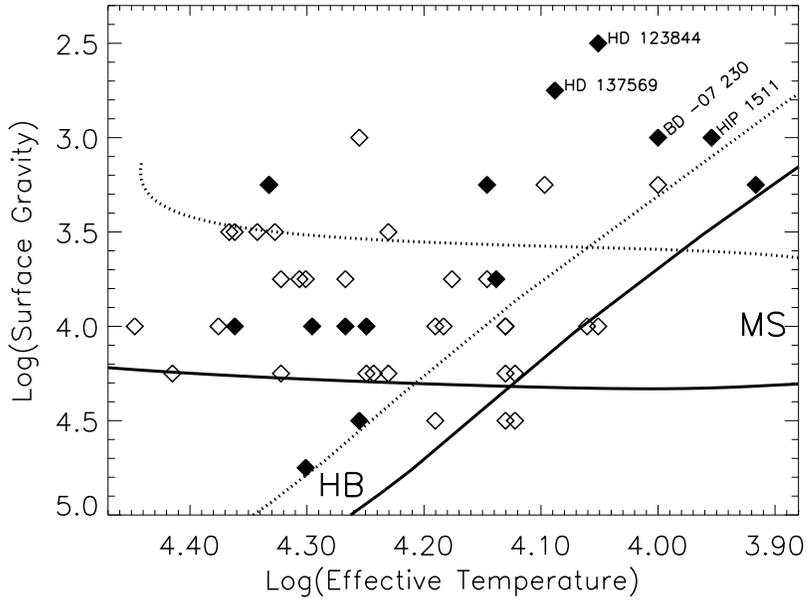}
\caption{Similar to Figure \ref{oestg}.  Solid diamonds: stars with
  vsin(i) $\le$ 40 km s$^{-1}$.  Open diamonds: stars with vsin(i) $\ge$
  40 km s$^{-1}$.} 
\end{figure}

\begin{figure}
\figurenum{10}
\label{hlbrotdist}
\includegraphics[angle=0,scale=1.]{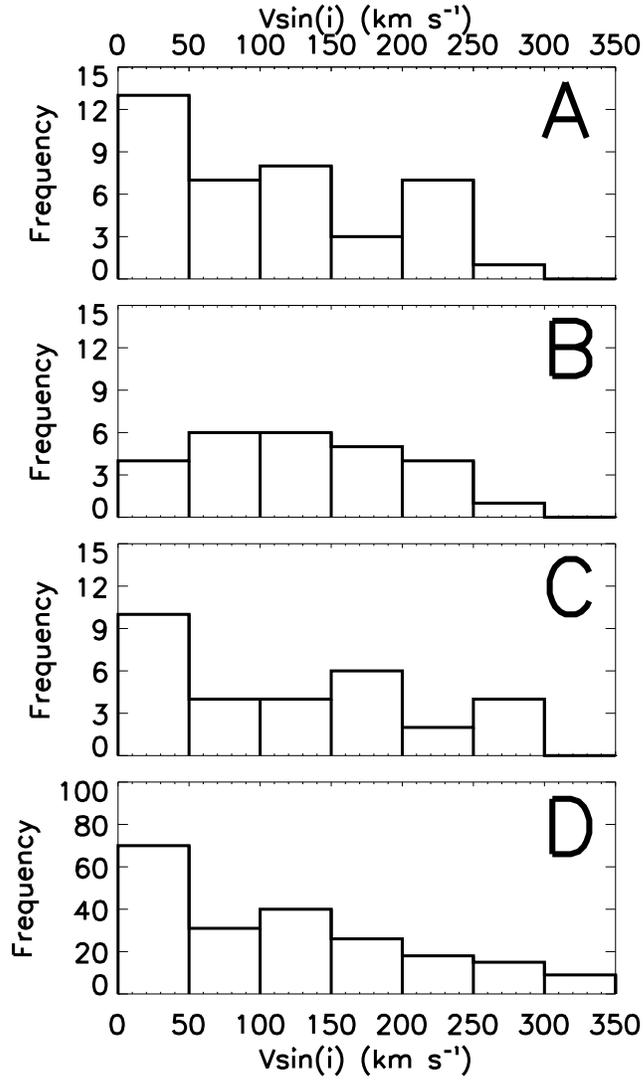}
\caption{Histograms of the vsin(i) distribution for various samples
  (bin size of 50 km~s$^{-1}$).  Panel A: all high latitude B
  stars in our sample; Panel B: only runaway stars in our sample.
  Panel C: high latitude B stars from \citet{magee98}; Panel D: B
  stars in the nearby field from \citet{wolff82}.}  
\end{figure}

The rotational velocity distribution for high latitude B stars is 
roughly the same as nearby B stars in the field (Figure
\ref{hlbrotdist}) with a slightly larger proportion of slow
rotating stars since these samples include BHB stars\citep{magee98}.
Other studies have biased their selection of runaway stars by using
stellar rotation as a criterion to distinguish between BHB stars and
``normal'' Population I stars (i.e. \citet{gs74} and \citet{behr03}).
But we have not.  There is no obvious correlation between the stars
with small vsin(i) and the main sequence or horizontal branch (Figure
\ref{slowrot}).  There are four slow vsin(i) stars clustered alone in
the upper left corner of that diagram.  However, three of those
(HD~123844, HD~137569, and HIP~1511) are clearly OES based on their
abundances and space velocities.  Because the runaway star sample is
selected without regard for projected rotational velocity (vsin(i)) we
are able to measure the unbiased rotational velocity distribution of
runaway B stars.  

Surprisingly, when the older evolved stars are removed from the
distribution (Figure \ref{hlbrotdist}, panel B) there is a clear
deficit of slow rotating stars compared to the field.  It is highly
unlikely that the field star samples of \citet{wolff82} and
\citet{guthrie84} are contaminated by slow rotating old evolved stars
because both are drawn from main sequence B stars in the Yale Bright
Star Catalog \citep{bsc}.  Furthermore, \citet{guthrie84} found that
the field stars have the same distribution as ``older''  OB
associations, which are not old enough to contain any post-main
sequence BHB or post-HB stars. 

\begin{figure}
\figurenum{11}
\label{guthrie}
\includegraphics[angle=0,scale=1.]{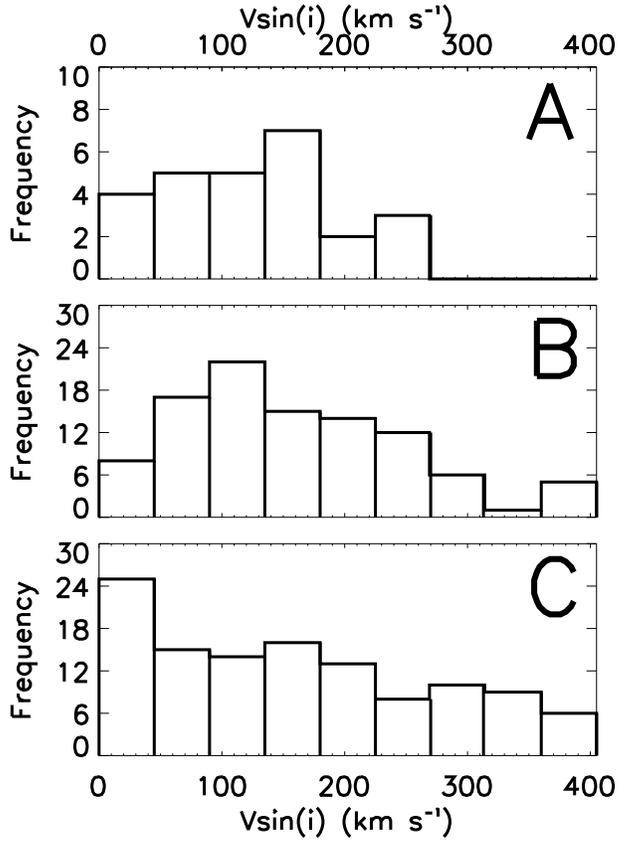}
\caption{Histograms of the vsin(i) distribution for various samples
  (bin size of 45 km~s$^{-1}$).  Panel A: runaway stars in our sample.
  Panel B: ``young'' OB associations from \citet{guthrie84}; Panel C:
  B stars in the nearby field from \citet{guthrie84}.} 
\end{figure}

Our runaway star distribution appears to be more like the ``youngest''
Population I B stars from \citet{guthrie84} (Figure \ref{guthrie}).  A
Kolmogorov-Smirnov test \citep{kennedy86} shows that there is a 50\%
probability that our runaway sample was drawn from Guthrie's
``young'' sub-group versus a 30\% chance that it was drawn from
Guthrie's field star sample\footnote{Adding one star to our runaways
  with a vsin(i) $\ge$ 180 km s$^{-1}$ increases the significance of
  the match with Guthrie's ``young'' sub-group to over 80\% with no
  change to the much lower probability of association with Guthrie's
  field sample.}. Guthrie's ``young'' OB associations were selected
for their high stellar density and \citet{strom05} and \citet{huang03}
have shown that denser star forming regions contain significantly
fewer slow rotating stars.  When one considers that dynamical ejection
(DES) is more efficient in denser stellar environments, the
implication is that most of the stars in our sample could have come
from these tightly packed OB associations.

\begin{figure}
\figurenum{12}
\label{rotvsvej}
\includegraphics[angle=90,scale=0.5]{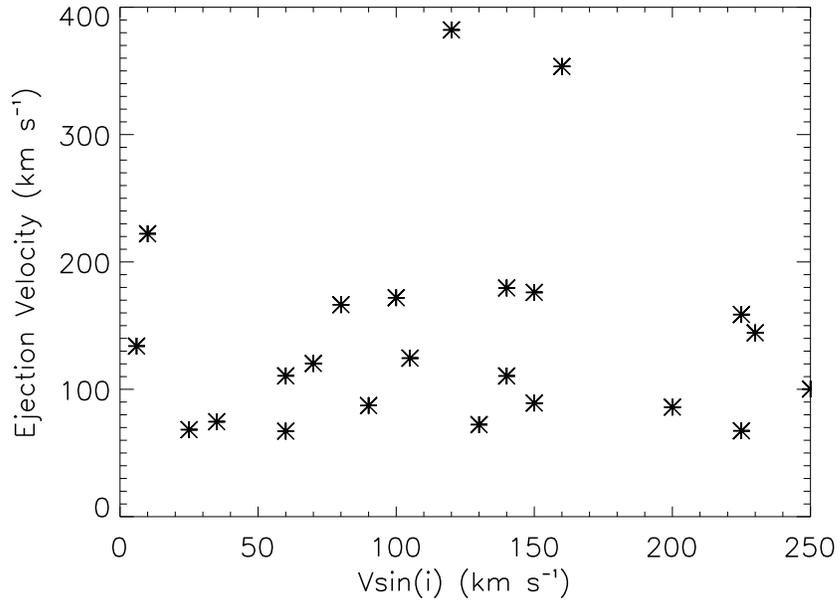}
\caption{Vsin(i) versus full ejection velocity (V$_{ej}$) for runaway
  stars in the sample} 
\end{figure}

The BSS mechanism involves the ejection of a secondary from a close
binary when the primary goes supernova.  In the case of a close binary
system there will be tidal locking, so that the rotation rate of
the runaway will be directly related to its original orbital velocity.  
It is possible that the combination of the bias in our sample (which
sharply excludes the smallest ejection velocities, see Section
\ref{runawaybias}) and pre-ejection BSS tidal locking could produce a
sample like ours with few slow rotators.  However if BSS is the
dominate mechanism at work in our sample, there should be a linear
relationship between ejection velocity and vsin(i) because the
ejection velocity should be comparable to the orbital velocity when
the system becomes unbound.  There is clearly no trace of any
relationship between rotation and ejection velocity in our sample
(Figure \ref{rotvsvej}).  Therefore, it seems unlikely that BSS is the
dominate mechanism which produced these runaways. 

On the other hand, DES is much more efficient in densely populated 
environments (like Guthrie's ``young'' OB associations) since it 
accelerates runaway stars through random dynamic interactions between 
stars or pairs of stars.  Because the distribution of vsin(i) for our 
sample so closely matches Guthrie's ``young'' sub-groups it seems 
likely that DES is the dominate mechanism in our runaway sample. 

\citet{hoogerwerf01} appears to contradict this notion by concluding 
that about two-thirds of massive runaways are produced by BSS.
However, they surveyed a subset of stars within 700 pc of the Sun with
space velocities up to 115 km s$^{-1}$ that are mostly excluded from our
sample.  It is possible that these are a sub-population with different
properties from the high latitude runaway B stars.  Their
classification scheme also assumed that BSS would produce runaways
with larger rotational velocities and with flight times shorter than
the estimated age of their ``parent groups.''  The former assumption
was made without the benefit of the work of \citet{strom05} or the
findings that we have reported here.  The later neglects that DES can
still occur in dense clusters as they age and expand, albeit at a
lower rate. 

\subsection{The Missing Runaway Be Stars}
If we assume that the frequency of Be stars is the same as reported by
\citet{zorec97} for the nearby field, this sample should have
at least ten Be stars out of the 48 high latitude B stars in
our sample.  However, there are no classical Be stars in our
sample\footnote{The spectrum of Feige 23 has narrow NaD emission lines
  (see Paper~I, Fig.~12) but no Hydrogen Balmer emission.}.  This
result is rather significant even considering that the number of Be
stars will always be under-counted by a survey of this type.  (Be
emission is a transient phenomenon that can be missed by surveys with
short temporal baselines.)

Two surveys have identified runaway Be stars at smaller distances from
the galactic plane.  \citet{slettebak97}
identified 8 runaway Be stars between 0.2--0.9 kpc from the plane
which barely overlaps the volume occupied by most of our sample and
contains mostly runaways that would have been ejected with V$_{ej}$
too low to be included in our sample.  Likewise, \citet{berger01}
found 3\%--7\% of the Be stars in the Hipparcos catalog have runaway
space velocities.  However those velocities ranged from 40 km s$^{-1}$
-- 102 km s$^{-1}$ which would have only a small overlap with our
sample.  Both these studies imply that the incidence of Be stars among
runaways is lower than measured in the field.  

A population of B stars which do not exhibit the Be phenomenon could
prove invaluable for identifying the critical factors involved.
Obviously, fast rotation alone does not create Be stars since on
average most runaways rotate faster than B stars found in the field
(see Section \ref{runawayrot}).  \citet{berger01} conjectured that
mass transfer during the post main sequence phase of a binary system
prior to BSS might play a role in the Be phenomenon.  There is a low
incidence among runaways of binaries that could transfer mass at
critical times (see Section \ref{binaries}).  However, it appears that
most of these stars were ejected by DES.  By process of elimination,
it remains that this could somehow be tied to the small number of
binaries or the origin of the runaways in dense star clusters with
some mechanism selecting against the Be phenomenon there. 

\citet{mcswain05} found no correlation between the
number of Be stars and stellar density in galactic star clusters.
But they did find a smaller frequency of Be stars (2\%--7\%) in clusters
than reported in the field by \citet{abt87} and \citet{zorec97}.
They give three reasons for this discrepancy:
\begin{enumerate}
\item{Magnitude limited field samples are biased in favor of early type B
stars in Gould's Belt and early spectral types have a higher Be
frequency.}
\item{Gould's Belt is also a population that has a higher frequency of
  Be stars due to its age.} 
\item{McSwain \& Gies made one-shot observations of their clusters
  which tend to underestimate the actual number of Be
  stars.} 
\end{enumerate}
However, the fact remains that \citet{berger01} also found a lower
incidence of Be stars among runaways than in the field.  Also like
\citet{mcswain05}, we do not have a long baseline of data which will
catch temporarily inactive Be stars.  If, as implied by the
vsin(i) distribution, the runaways are ejected mainly from cluster
environments and we assume Be stars are produced in the smaller
proportions measured by \citet{mcswain05} or \citet{berger01}, then it
is much more likely that our sample of 49 stars with 31 Population I
runaways would not contain any Be stars.

\section{Properties of the Old Evolved Stars\label{discuss2}}

\subsection{Revised Kinematics of BHB Stars\label{bhbvel}}
The distances and space velocities of the low mass evolved stars in
the sample were initially exaggerated by estimating their photometric
parallax from a main sequence luminosity (see Section \ref{sublum}).
Now that the BHB stars are identified, their photometric parallax and
space velocities can be recalculated using more appropriate absolute
magnitudes from the evolutionary tracks of \citet{dorman93} for
the appropriate effective temperatures, surface gravities, and
metallicities (Table \ref{bhbkin}) and the bolometric corrections
calculated by ATLAS12.  HD~21305 has a very small proper
motion so a factor of five change in its distance has very little
effect on its space velocity.  However, the space velocities of
HIP~1511, HIP~41979, PB~166, and HD~222040 are markedly different.
The revised space velocities for these stars are much more reasonable
than those listed in Table \ref{resultstab}.

It is easy to chemically and kinematically assign HIP~1511 and
HD~222040 membership in the halo population and likewise assign
HD~21305 membership in the thick disk population.  HIP~41979 and
PB~166 have abundances which are consistent with the thick disk but
their revised V$_{\pi}$ and V$_z$ do not comfortably fit in the
expected velocity distribution for the thick disk population (Table
\ref{tab61}).  The uncertainty of the distances and the proper motions
have significant influence over V$_{\pi}$ and V$_z$ for both
stars.  Based only on V$_{\theta}$ it seems likely that both are
members of the thick disk population.  

We also recalculated the distances and space velocities of the post-HB
and ``ambiguous'' evolved stars using the Dorman et al. tracks (Table
\ref{bhbkin}).  Because there is a chance that the stars labeled
``ambiguous'' could be main sequence stars, we have included their
main sequence distances and velocities for comparison.  In the case of
non-BHB stars we are less certain that the tracks give the true
luminosities because we are unsure if they match that evolutionary
state.  The Dorman et al. tracks yield distances of 2.4 kpc and 0.83
kpc for BD~+33~2642 and BD~+13~3224 respectively.  However, those
values are significantly different from the distances obtained
independently by \citet{napiwotzki94} (3.30 kpc) and \citet{jeffery01}
(1.70 kpc).  We have used those more reliably determined distances
from the literature to calculate the space velocities for those stars.
In any case, when we use these distances to calculate the space
velocities of the post-HB and ``ambiguous'' stars all of them appear
to be members of the galactic thick disk population except
BD~+33~2642, which is probably a member of the halo population. 

\subsection{Metal-Rich BHB Stars\label{bhbmetal}}

In Paper~I, we identified three metal-rich BHB stars from their
abundance patterns which are heavily influenced by atomic diffusion:
HD~21305, HIP~41979, and PB~166.  The analysis of the space velocities
of HIP~41979 and PB~166 confirm they are BHB stars.  In the case of
HD~21305, its effective temperature and gravity are not consistent
with a horizontal branch star.  The space velocity of HD~21305
indicates that it is probably a member of the galactic thick disk
population regardless of its evolutionary state (see Section
\ref{bhbvel}).  Therefore, its BHB classification is made entirely
based on its abundance pattern and hence, is less certain than the
other two metal-rich BHB stars.  

\section{Conclusions}
We have sorted the high galactic latitude stars in our
sample into five groups:  Population I runaways (31 stars), BHB stars (5
stars), post-HB stars (3 stars), old evolved stars of ambiguous
classification (6 stars), and those that we cannot classify (2
stars).  The two unclassified stars are probably extreme
Population I runaways.  

We agree with other studies that roughly two thirds of faint high
galactic latitude B stars are Population I runaways ejected from the galactic
disk.  The remainder are older evolved post-main sequence stars.  No
star in this sample (or any other we are aware of) shows metal poor
abundances and halo-like kinematics of a young massive star which
formed in situ in the galactic halo. HD~237840 and HD~233622 are
classified as old evolved stars based on their abundances.  However
they are rotating far too rapidly to be low mass stars.  They could be
chemically odd Population I runaways or main sequence stars formed in
situ in the halo but there is not enough evidence to justify either
alternative. 

The Population I runaways have a number of interesting properties:
\begin{itemize}
\item{A low incidence of binaries (less than 10\% as shown in other
  studies).}
\item{A deficit of slow rotators (vsin(i) $<$ 50 km s$^{-1}$) compared
  to the field.} 
\item{No Be stars.}
\end{itemize}

No significant shift was measured in the radial velocities of the 31
runaways in the sample over baselines of a few days or months except
for HD~137569 and the eclipsing binary HD~138503 (IT Lib).  That binary
was probably launched by DES since it is an SB2 comprised of two
normal main sequence stars.  Two stars
(HD~188618 and HD~216135) have discrepancies between their measured
radial velocities and those reported in the literature.  These could
be single-lined spectroscopic binaries (SB1s) with neutron star
companions produced by BSS with periods on the order of several
hundred days. These results are not significant by themselves but they
are consistent with more extensive studies which show that the binary
frequency among early type runaways is less than 10\%.  This relative
lack of runaway binaries implies that DES is the dominant mechanism.
The projected rotational velocity (vsin(i)) distribution also supports
this hypothesis. 

The rotation of Population I runaways are like densely populated star
clusters in that they have fewer slow rotators than the field population.
Therefore, we conclude that most of the runaways originate in dense
clusters or star forming regions.  This supports the DES hypothesis
since it is much more efficient in dense stellar environments.  There
is also no correlation between ejection velocity (V$_{ej}$) and
vsin(i), as would be expected for a BSS dominated sample.

Together, the low frequency of binaries and the lack of slow rotators
provide a compelling argument that DES is the dominant ejection
mechanism for runaway B stars.  This is not to say that BSS does not
contribute at some level.  HD~188618 and HD~215135 could be long
period SB1s of the type produced by BSS.  However, it appears that
runaways accelerated by DES dominate this sample.

The absence of Be stars from the sample presents another interesting
facet.  This could be due to an observational bias either
on our part or affecting field star samples.  But it could also
potentially be due to some mechanism responsible for the Be phenomenon
that runaways do not possess.  The lack of binary companions,
something about the ejection mechanism, or another factor in the
cluster environment which they were ejected from may leave
runaways without crucial ingredients necessary to become Be stars.
Further study of runaways could help unlock the mystery of the Be
phenomenon. 

The old evolved stars also exhibit interesting characteristics.  The
velocities of the BHB stars determine their membership in either the
thick disk or halo populations.  We confirmed that PB~166 and
HIP~41979 are nearby solar-metallicity BHB stars which belong to the
thick disk.  There is still some uncertainty about HD~21305 which has
a pattern of abundances like the other two and kinematically belongs
to the thick disk.  However its log(g) and T$_{eff}$ place it well
away from the TAHB. 

\section{Acknowledgments}
I wish to extend hearty thanks to R.E. Luck for his support of
this work as my PhD adviser and his input on this publication.  Thank
you also to the other members of my PhD committee for their critical
comments on my dissertation:  H. Morrison, J.C. Mihos, and R. Dunbar.
I also thank P. Harding for generously sharing his galactic potential
integrator and V. McSwain and S. Wolff for their very useful comments
and discussion concerning Be star populations, star clusters, and
stellar rotation.  Thank you also to the anonymous referee whose
critical comments helped polish this publication.  I feel this
work was strengthened by their diligence and willingness to
give significant time and effort without credit to their name.  Last
but not least, I want to thank Kris Davidson and Roberta Humphreys for
providing me with the time, environment, and encouragement to finish
distilling this research into a publishable form. This work was
supported in part by the Jason J. Nassau Scholarship Fund and the
Townsend Fund through the generous continued support of the Ford,
Nassau, and Townsend families. 





\clearpage

\appendix
\section{Computing Trajectories in the Galactic Potential\label{galmod}\label{ApB}}
The orbital integrator we use is described in detail by
\citet{harding01}. 
The Milky Way's potential was simulated following the prescription of
\citet{johnston95} using a three component model which combines:  (1)
a \citet{miyamoto75} disk potential, (2) a \citet{hernquist90}
spheroidal potential, (3) and a logarithmic dark halo potential.  The
parameters for each of the components are selected so that together
they reproduce the observed rotation curve of the Milky Way's disk
(Table \ref{tab66}).  \citet{harding01}, \citet{johnston02},
\citet{helmi01} have used the same model and parameters to
successfully simulate the motions of stars in the galactic halo.
This model and integrator also reasonably reproduce the
results of \citet{magee01} from their data (Table \ref{tab67}) within the 
margin of error we would expect since they use a different galactic 
potential model.  

The parameters of the galactic potential model 
affect the flight times of runaways from the galactic plane.  The
disk component of the model provides most of the restoring force that
governs the ballistic trajectories calculated for each star.
Therefore, it is a matter of concern that the mass surface density of
the galactic disk we have adopted in our model for this study
($\Sigma_{disk}$ = 95 M$_{\sun}$ pc$^{-2}$) is more than twice the
local mass surface density of the disk alone ($\Sigma_{disk}$ = 48$\pm$9
M$_{\sun}$ pc$^{-2}$, \citet{kuijken89};  $\Sigma_{disk}$ = 56$\pm$6
M$_{\sun}$ pc$^{-2}$, \citet{holmberg04}) and about four thirds the
total integrated mass density within 1.1 kpc of the galactic plane
($\Sigma_{disk}$ = 71$\pm$6 M$_{\sun}$ pc$^{-2}$,\citet{kuijken91};
$\Sigma_{disk}$ = 74$\pm$6 M$_{\sun}$ pc$^{-2}$,\citet{holmberg04}). 

In simple terms, a decrease in $\Sigma_{disk}$ should produce a
proportional increase in the flight times of the stars in the sample.
A smaller $\Sigma_{disk}$ translates into a smaller restoring force
for stars displaced from the galactic plane, allowing runaway stars
with a given ejection velocity to travel further from the disk.  This
means that stars would travel at lower velocities to reach the same
positions, lengthening their flight times.  The disk potential is
embedded in the larger halo potential, so stars that journey beyond
the immediate influence of the disk would be less effected by changes
in $\Sigma_{disk}$.  For this reason, this effect should be greater
for stars nearer to the disk moving with slower initial ejection
velocities than for stars with higher initial ejection velocities
further from the plane. 

The adopted potential model (Table \ref{tab66}) relies on a
balance between its three components in order to reproduce the
observed rotation curve of the galactic disk.  Therefore, a complete
reformulation of the model is beyond the scope of this study.
However, a simple reassessment gives an reliable estimate of the
effect that a lower $\Sigma_{disk}$ might have on the calculated
flight times.  In a simple-minded way, we lowered $\Sigma_{disk}$ by
altering the parameters for the disk component and then adjusted the
bulge and halo components to flatten the galactic rotation curve as
near to 210 km s$^{-1}$ as possible between 5 kpc and and 20 kpc from
the galactic center.  In three of the models $\Sigma_{disk}$ was
altered by changing the total mass of the disk ($M_{disk}$).  However
a fourth model label "Short Disk" was produced by adopting a shorter
disk scale length favored by \citet{kent91}, \citet{spergel96},
\citet{drimmel01}, and references therein.  The parameters for each
model are given in Table \ref{tab69}.

\begin{figure}
\figurenum{13}
\label{difmods}
\plotone{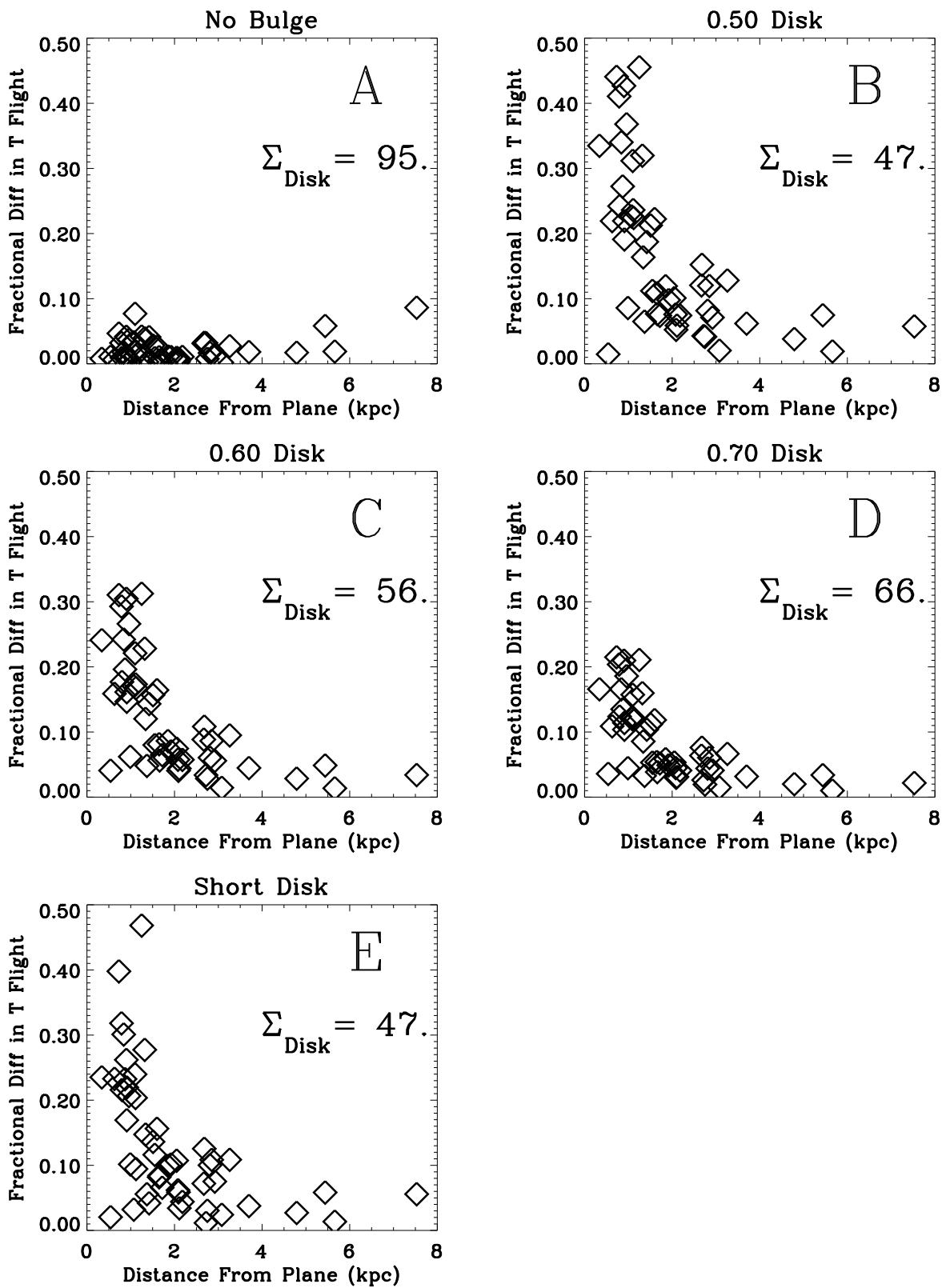}
\caption{Panels A through E show the fractional difference of the
  flight times for each model described in Table \ref{tab69} relative
  to the baseline model as a function of the star's distance from the
  galactic plane.}
\end{figure}

The flight times were recalculated for each star in the sample using
the alternate models.  Figure \ref{difmods} shows the effect of each
model on the calculated flight times versus distance of the star from
the galactic plane.  The No Bulge model is identical to the original
model except that it has no bulge component.  Panel A confirms that
the bulge component has no significant effect on the calculated flight
times.  The results from the Short Disk model (panel E) are similar to
the 0.50 Disk model (panel B) which has the same $\Sigma_{disk}$.
This leads us to conclude that the flight times are not very sensitive
to the disk scale length except through its influence on
$\Sigma_{disk}$. 

Panels B through D demonstrate that reducing $\Sigma_{disk}$ lengthens
the flight times for stars in the sample.  The transition between
stars dominated by the disk and the halo occurs at about 1.5 kpc from
the galactic plane.  The flight times for stars more than 1.5 kpc from
the galactic plane are at most ten to fifteen percent larger while
stars that are within 1.5 kpc of the disk have much larger
differences in calculated flight times dependent on $\Sigma_{disk}$.   

\begin{figure}
\figurenum{14}
\label{msvstd2}
\includegraphics[angle=90,scale=0.5]{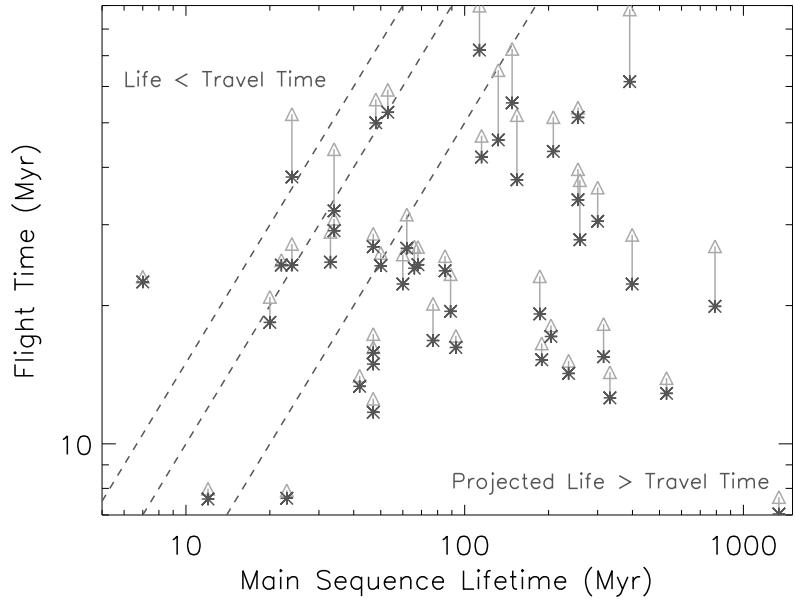}
\caption{The flight time of the star versus its estimated main
  sequence lifetime.  An asterisk (*) denotes the flight time
  calculated with the galactic potential model parameters in Table
  \ref{tab66}.  The triangles denote the flight time
  calculated using the 0.5 Disk model (Table \ref{tab69}).}
\end{figure}

The differences in travel time between the original model and the 0.5
Disk model (B) affect the relationship between each star's flight time
and main sequence lifetime (Figure \ref{msvstd2}).  Fortunately, most
stars remain within the same region defined by the dotted lines if
$\Sigma_{disk}$ is halved.  Therefore, we conclude that changes to the
parameters of the galactic potential within the range of values
discussed by the literature have no affect on the classification of
stars in our sample.

\clearpage

\clearpage

\begin{deluxetable}{lrrrr}
\tablewidth{0pt}
\tabletypesize{\tiny}
\tablecolumns{5}
\tablecaption{Kinematic Parameters of Galactic Stellar Populations\label{tab61}}
\tablehead{
 &\colhead{V$_{\theta}$}&\colhead{$\sigma_{\pi}$}&\colhead{$\sigma_{\theta}$}&\colhead{$\sigma_{z}$}\\
\colhead{Population}&\colhead{(km s$^{-1}$)}&\colhead{(km s$^{-1}$)}&\colhead{(km s$^{-1}$)}&\colhead{(km s$^{-1}$)}}
\startdata 
Thin Disk\tablenotemark{a}&214&34&21&18\\
Thick Disk\tablenotemark{a}&184&61&58&39\\
Halo\tablenotemark{b}&25&135&105&90\\
\enddata
\tablenotetext{a}{\citet{edvardsson93}}
\tablenotetext{b}{\citet{binneymerrifield98}}
\end{deluxetable}

\begin{deluxetable}{lrrlrrrrrrr}
\tablewidth{0pt}
\tabletypesize{\tiny}
\tablecolumns{11}
\tablecaption{Parameters of Sample Stars\label{inputtab}}
\tablehead{
 &\multicolumn{3}{c}{Average Proper Motion}&\multicolumn{2}{c}{Radial  Vel\tablenotemark{b}}&\colhead{Apparent}&&\colhead{MS Mass\tablenotemark{d}}&\colhead{Absolute}&\colhead{Distance}\\
\colhead{Star}&\colhead{$\mu_{\alpha}$ (mas yr$^{-1}$)}&\colhead{$\mu_{\delta}$ (mas  yr$^{-1}$)}&\colhead{Source\tablenotemark{a}}&\colhead{km s$^{-1}$}&\colhead{N}&\colhead{V  mag}&\colhead{A$_{V}$\tablenotemark{c}}&\colhead{(M$_\sun$)}&\colhead{V mag\tablenotemark{d}}&\colhead{(kpc)}
}
\startdata
HD~1112&57.0$\pm$0.5&10.6$\pm$0.5&ACT,CMC&-0.3$\pm$5.1&3&8.98&0.12&2.1&+0.63&0.44\\
HIP 1511&-3.8$\pm$1.4&-34.4$\pm$1.1&CMC&-229.4$\pm$2.4&40&11.61&0.2&1.7&-1.12&3.20\\
BD -15 115&8.0$\pm$1.0&0.1$\pm$0.6&ACT,UCAC1  &91.3$\pm$5.7&64&10.87&0.06&6.8&-1.73&3.22\\
HD~8323&-4.9$\pm$0.6&-17.7$\pm$0.4&ACT,UCAC1  &-7.0$\pm$2.4\tablenotemark{g}&\nodata&9.52&0.13&2.7&+0.02&0.75\\
BD -7 230&13.2$\pm$0.6&-1.2$\pm$0.8&ACT,UCAC1  &-2.3$\pm$1.4&77&11.14&0.11&3.5&-2.12&4.27\\
Feige 23&-17.4$\pm$1.5&-12.9$\pm$1.5&CMC&0&1&11.08&0.09&2.4&+0.80&1.09\\
HD~15910&4.8$\pm$0.9&-2.4$\pm$0.9&ACT,UCAC1  &-44.8$\pm$6.0&5&10.1&0.07&2.9&+0.32&0.87\\
HIP 12320&4.9$\pm$1.7&-2.4$\pm$1.6&CMC&24.1$\pm$12.1&2&12.04&0.27&3.7&-0.46&2.80\\
HD~21305&-1.4$\pm$0.7&2.1$\pm$0.8&ACT,UCAC1  &73.5$\pm$6.7&48&10.35&0.03&6.1&-1.59&2.41\\
HD~21532&8.6$\pm$0.7&3.2$\pm$0.5&ACT,UCAC1,CMC&27.8$\pm$6.4&25&9.94&0&3.4&+0.31&0.84\\
HD~40267&-4.0$\pm$0.7&8.3$\pm$0.6&ACT,UCAC1,CMC&53.9$\pm$3.1&27&9.83&0.13&6.3&-2.71&3.06\\
BD +61 996&2.0$\pm$1.0&-5.9$\pm$0.8&ACT  &83.6$\pm$10.4&53&9.9&0.15&8.1&-4.28&6.57\\
HIP 41979&11.7$\pm$0.8&-23.4$\pm$0.7&ACT  &-2.8$\pm$2.7&102&11.48&0.11&6.0&+0.05&1.77\\
HD~233662&2.1$\pm$0.7&-9.8$\pm$0.7&ACT  &36.3$\pm$13.6&4&10.01&0.05&6.9&-3.06&4.01\\
HD~237844&3.5$\pm$1.0&-5.5$\pm$0.6&ACT  &13.6$\pm$11.8&12&10.11&0.02&5.9&-2.16&2.94\\
BD +16 2114&-12.3$\pm$0.7&4.3$\pm$0.6&ACT  &49.0$\pm$2.4\tablenotemark{g}&&9.98&0.14&3.7&-0.12&0.98\\
BD +38 2182&-6.7$\pm$1.3&0.1$\pm$0.8&ACT,CMC&84.2$\pm$10.5&7&11.25&0.06&6.6&-2.12&4.60\\
Feige 40&-1.5$\pm$0.7&-6.1$\pm$0.7&ACT,CMC&71.5$\pm$5.7&7&11.19&0.13&3.9&+0.56&1.26\\
HD~100340&3.6$\pm$1.1&11.4$\pm$0.9&ACT  &254.4$\pm$9.1&9&10.12&0.08&10.&-2.06&2.62\\
HD~103376&8.7$\pm$1.0&5.3$\pm$0.7&ACT  &20.0$\pm$1.6&8&10.22&0.03&3.0&+1.11&0.65\\
HD~105183&-8.3$\pm$0.9&-6.0$\pm$0.7&ACT,CMC&34.6$\pm$5.4&12&11.07&0.12&4.7&-1.47&3.05\\
HD~106929&-10.4$\pm$0.7&-2.5$\pm$0.7&ACT  &28.3&1&9.84&0.07&3.1&+1.13&0.53\\
BD +49 2137&-12.1$\pm$0.9&4.8$\pm$0.7&ACT  &115$\pm$10\tablenotemark{h}&\nodata&10.66&0.04&3.4&-0.45&1.64\\
BD +36 2268&-1.3$\pm$1.2&1.4$\pm$0.9&ACT  &29.2$\pm$2.4&16&10.48&0.03&6.9&-2.49&3.86\\
HD~110166&0.0$\pm$0.6&-5.0$\pm$0.6&ACT,CMC&-48.4$\pm$13.2&2&8.79&0.04&4.2&-2.17&1.53\\
BD +30 2355&-4.2$\pm$0.8&-8.4$\pm$1.0&ACT  &-5.1$\pm$7.7&8&10.64&0.08&3.2&-1.49&2.56\\
PB 166&-26.4$\pm$1.5&10.2$\pm$1.4&CMC&43.1$\pm$1.5&16&12.45&0.03&5.4&+0.39&2.54\\
Feige 84&-7.6$\pm$1.1&-8.7$\pm$0.8&ACT  &147.9$\pm$9.9&12&11.86&0.06&5.1&-0.65&3.09\\
HD~121968&1.4$\pm$0.9&17.8$\pm$0.7&ACT  &28$\pm$2.4\tablenotemark{g}&\nodata&10.26&0.2&7.4&-2.08&2.67\\
HD~123884&-0.4$\pm$0.8&-2.6$\pm$0.7&ACT,UCAC1  &16.6$\pm$2.6&17&9.36&0.27&3.2&-3.55&3.39\\
HD~125924&4.6$\pm$0.9&-11.7$\pm$0.7&ACT,CMC&239.0$\pm$ 2.4\tablenotemark{g}&\nodata&9.68&0.15&6.9&-1.95&1.98\\
BD +20 3004&-9.9$\pm$0.7&5.7$\pm$0.6&ACT  &27$\pm$10\tablenotemark{h}&\nodata&10.15&0.08&3.6&-2.21&2.85\\
HD~137569&8.5$\pm$0.5&-10.5$\pm$0.4&ACT,CMC&-45$\pm$3\tablenotemark{f}&\nodata&7.95&0.12&4.7&-3.50&1.84\\
HD~138503&1.8$\pm$0.7&-1.3$\pm$0.8&ACT,UCAC1  &-53.6$\pm$1.2\tablenotemark{e}&\nodata&9.1&0.7&8.1&-3.40\tablenotemark{e}&2.29\\
HD~140543&-3.4$\pm$0.7&-2.4$\pm$0.6&ACT,UCAC1  &-9$\pm$5\tablenotemark{h}&\nodata&8.88&0.84&22&-6.25&7.22\\
BD +33 2642&-12.7$\pm$1.1&1.6$\pm$1.0&ACT  &-92.3$\pm$ 2.1&26&10.84&0.04&10.&-4.25&10.24\\
HD~146813&0.2$\pm$0.7&-3.8$\pm$0.6&ACT  &19.1$\pm$5.7&34&9.06&0.03&8.2&-2.58&2.10\\
HD~149363&-7.6$\pm$0.9&-12.8$\pm$0.6&ACT  &145.8$\pm$7.1&49&7.79&0.87&15.&-3.49&1.21\\
BD +13 3224&-4.6$\pm$1.0&3.3$\pm$0.6&ACT &2.42$\pm$0.53\tablenotemark{i}&\nodata&10.53&0.18&6.9&-1.94&2.88\\
HD~183899&1.7$\pm$0.7&-0.2$\pm$0.6&ACT,UCAC1,CMC&-51.3$\pm$6.0&28&11.&0.56&9.5&-3.83&4.11\\
HD~188618&-3.4$\pm$0.7&3.5$\pm$0.4&ACT,UCAC1  &29.0$\pm$5.6&12&9.38&0.79&9.7&-3.67&2.83\\
AG +03 2773&4.2$\pm$0.8&-7.3$\pm$0.7&ACT  &15.6$\pm$13.2&24&10.51&0.21&1.6&+2.44&0.40\\
HD~206144&-6.4$\pm$0.8&-11.0$\pm$0.6&ACT,UCAC1,CMC&116.7$\pm$8.1&21&9.33&0.24&9.7&-3.59&3.52\\
HD~209684&0.7$\pm$0.7&-0.5$\pm$0.5&ACT,UCAC1  &71.6$\pm$7.7&28&9.86&0.12&6.8&-1.23&1.56\\
HD~213781&-0.5$\pm$0.6&3.6$\pm$0.5&ACT,UCAC1,CMC&-32.7$\pm$6.1&17&9.04&0.19&4.1&-1.18&1.01\\
HD~216135&-18.5$\pm$0.7&-13.8$\pm$0.6&ACT,UCAC1  &14.5$\pm$4.1&11&10.13&0.14&4.5&-0.89&1.49\\
HD~218970&46.5$\pm$0.8&33.4$\pm$0.7&ACT,UCAC1,CMC&93.8$\pm$4.5&11&9.75&0.08&5.2&-0.71&1.19\\
HD~220787&3.5$\pm$0.7&-1.4$\pm$0.5&ACT,UCAC1,CMC&26.4$\pm$2.6&18&8.3&0.09&5.7&-1.35&0.82\\
HD~222040&23.6$\pm$1.2&-26.9$\pm$0.9&ACT,UCAC1&75.8$\pm$1.5&22&10.87&0.08&1.1&+0.23&1.30\\
\enddata
\tablenotetext{a}{Sources averaged into the proper motion in addition to HIP and Tycho-2.}
\tablenotetext{b}{Unless otherwise noted the radial velocity is measured from the spectra.  N is the number of  independent line measures that went into the radial velocity}
\tablenotetext{c}{See Paper~I for details on A$_{V}$.}
\tablenotetext{d}{Absolute magnitude obtained from \citet{schaller92}
  using T$_eff$, log(g), and [Fe/H] assuming that the star is on the
  hydrogen burning main sequence with bolometric correction from
  ATLAS12 \citep{kurucz}.} 
\tablenotetext{e}{Radial velocity and absolute magnitude from \citet{itlib}.}
\tablenotetext{f}{Radial velocity from \citet{bolton80}.}
\tablenotetext{g}{Radial velocity from \citet{gs74}.}
\tablenotetext{h}{Radial velocity from \citet{evans79}.}
\tablenotetext{i}{Radial velocity from \citet{lynasgray84}.}
\end{deluxetable}

\begin{deluxetable}{lrlrrrr}
\tablewidth{0pt}
\tabletypesize{\tiny}
\tablecolumns{7}
\tablecaption{Radial Velocity Standards\label{stdrvtab}}
\tablehead{
&\colhead{Apparent}&\colhead{Spectral}&\colhead{vsin($i$)\tablenotemark{a}}&\colhead{Standard  RV\tablenotemark{a}}&\colhead{Measured RV}\\
\colhead{Star}&\colhead{V Mag}&\colhead{Class}&\colhead{km s$^{-1}$}&\colhead{km s$^{-1}$}&\colhead{km  s$^{-1}$}&\colhead{N\tablenotemark{b}}
}
\startdata
HR 675&5.03&A2 V&18&0.43$\pm$0.05&0.3$\pm$0.5&277\\
HR 1389&4.29&A2 IV-V&8&38.97$\pm$0.12&38.9$\pm$1.0&75\\
HR 1397&6.06&B6 IV&25&12.04$\pm$1.00&11.8$\pm$1.5&151\\
HR 2010&4.91&B9.5 V&21&20.55$\pm$0.12&19.3$\pm$1.2&235\\
HR 2238&4.48&A1 V&41&-2.01$\pm$0.95&-2.1$\pm$0.1&426\\
HR 2489&6.46&A9 III&8&7.68$\pm$0.07&8.9$\pm$0.7&68\\
HR 2818&4.64&A0 IV&14&26.81$\pm$0.15&27.4$\pm$0.4&527\\
HR 3136&5.65&A1 V&29&44.22$\pm$0.26&44.1$\pm$0.1&492\\
HR 3383&5.81&A1 V&4&2.79$\pm$0.09&3.7$\pm$0.3&393\\
HR 4033&3.45&A1 IV&44&18.08$\pm$1.03&18.9$\pm$0.8&61\\
HR 4187&5.80&A0 V&14&-15.45$\pm$0.93&-17$\pm$1.3&61\\
HR 4359&3.34&A2 IV&18&7.44$\pm$0.16&7.6$\pm$0.5&276\\
HR 7773&4.76&B9.5 V&18&-1.02$\pm$0.24&-1.1$\pm$0.7&435\\
HR 8404&5.80&B9.5 V&4&0.20$\pm$0.05&-0.1$\pm$0.3&615\\
HR 8768&6.39&B2 V&10&-11.37$\pm$0.80&-10.8$\pm$1.8&52\\
\enddata
\tablenotetext{a}{Standard radial velocity and Vsin($i$) taken from \citet{fekel99}.}
\tablenotetext{b}{N is the number of independent line measures averaged in the measured radial velocity.}
\end{deluxetable}

\begin{deluxetable}{lrrl}
\tablewidth{0pt}
\tabletypesize{\tiny}
\tablecolumns{4}
\tablecaption{Radial Velocities Versus Published Values\label{rvcomp}}
\tablehead{
&\multicolumn{2}{c}{Radial Velocity (km s$^{-1}$)}\\
\colhead{Star}&\colhead{This Work}&\colhead{Literature}&\colhead{Reference}
}
\startdata 
HD 1112&-0.3$\pm$5.1&6.0$\pm$2.4&\citet{gs74}\\
\nodata&\nodata&-9.4$\pm$9.7&\citet{behr03}\\
BD -15 115&91.3$\pm$5.7&86.7&\citet{graham73}\\
\nodata&\nodata&93.0$\pm$4.0&\citet{ramspeck01b}\\
\nodata&\nodata&94.0&\citet{conlon92}\\
BD -7 230&-2.3$\pm$1.4&2.0$\pm$2.4&\citet{gs74}\\
\nodata&\nodata&-1.7$\pm$0.8&\citet{behr03}\\
Feige 23&0.0$\pm$10.0\tablenotemark{a}&6.0$\pm$2.4&\citet{gs74}\\
HD 15910&-44.8$\pm$6.0\tablenotemark{b}&-79.0$\pm$13.0&\citet{harding71}\\
HIP 12320&24.1$\pm$12.1\tablenotemark{a}&-5.0$\pm$22.0&\citet{gs74}\\
HD 21532&27.8$\pm$6.4&22.0$\pm$2.4&\citet{gs74}\\
HD 233662&36.3$\pm$13.6\tablenotemark{b}&11.0&\citet{duflot95}\\
\nodata&\nodata&31.8$\pm$22.3&\citet{behr03}\\
BD +38 2182&84.2$\pm$10.5&79.0$\pm$22.0&\citet{gs74}\\
Feige 40&71.6$\pm$5.7&130.0$\pm$26.0&\citet{gs74}\\
\nodata&\nodata&74.2$\pm$14.0&\citet{behr03}\\
HD 100340&254.4$\pm$9.1&248.0$\pm$22.0&\citet{gs74}\\
\nodata&\nodata&250.0&\citet{keenan87}\\
\nodata&\nodata&246.0$\pm$5.0&\citet{ryans99}\\
\nodata&\nodata&256.9$\pm$19.9&\citet{behr03}\\
HD 103376&24.1$\pm$1.6&18.0&\citet{duflot95}\\
\nodata&\nodata&7.9$\pm$17.6&\citet{behr03}\\
HD 105183&34.6$\pm$5.4&34.0$\pm$2.4&\citet{gs74}\\
\nodata&\nodata&22.0&\citet{dufton93}\\
\nodata&\nodata&32.6$\pm$15.3&\citet{behr03}\\
HD 106929&28.3$\pm$10.0\tablenotemark{a}&16.6$\pm$3.7&\citet{grenier99}\\
BD +36 2268&29.2$\pm$2.4&31.0&\citet{duflot95}\\
\nodata&\nodata&41.0$\pm$5.9&\citet{behr03}\\
HD 110166&-48.4$\pm$13.2\tablenotemark{b}&-89.9$\pm$2.6&\citet{barbier00}\\
BD +30 2355&-5.1$\pm$7.7&-12.0$\pm$2.4&\citet{gs74}\\
\nodata&\nodata&1.2$\pm$25.0&\citet{behr03}\\
PB 166&43.1$\pm$1.5&60.0&\citet{conlon89a}\\
\nodata&\nodata&43.0&\citet{deboer88}\\
Feige 84&147.9$\pm$9.9&148.0$\pm$26.0&\citet{gs74}\\
\nodata&\nodata&153.0$\pm$27.8&\citet{behr03}\\
HD 123884&16.6$\pm$2.6&6.6&\citet{duflot95}\\
BD +33 2643&-92.3$\pm$2.1&-100.0$\pm$2.4&\citet{gs74}\\
\nodata&\nodata&-94.0&\citet{napiwotzki01}\\
\nodata&\nodata&-94.0$\pm$2.5&\citet{behr03}\\
HD 146813&19.1$\pm$5.7&21.0$\pm$5.0&\citet{grenier99}\\
HD 149363&145.8$\pm$7.2&141.4$\pm$2.0&\citet{barbier00}\\
HD 183899&-51.3$\pm$6.0&-45.0&\citet{duflot95}\\
HD 188618&29.0$\pm$5.6&-15.0&\citet{duflot95}\\
HD 206144&116.7$\pm$8.1&112.0$\pm$11.3&\citet{kilkenny75}\\
HD 213781&-32.7$\pm$6.1&-31.0$\pm$2.2&\citet{gs74}\\
\nodata&\nodata&-30.0$\pm$3.5&\citet{behr03}\\
HD 216135&14.5$\pm$4.1&-17.0$\pm$2.4&\citet{gs74}\\
HD 218970&93.8$\pm$4.5&98.0$\pm$2.4&\citet{gs74}\\
HD 220787&26.4$\pm$2.6&24.9$\pm$1.5&\citet{barbier00}\\
\nodata&\nodata&26.5$\pm$2.4&\citet{behr03}\\
HD 222040&75.8$\pm$1.4&65.0$\pm$9.0&\citet{harding71}\\
\enddata
\tablenotetext{a}{Only Hydrogen Balmer lines were used to measure this radial velocity.}
\tablenotetext{b}{Only Hydrogen Balmer and He~I lines were used to measure this radial velocity.}
\end{deluxetable}

\begin{deluxetable}{lrrrrrrrrrr}
\tablewidth{0pt}
\tabletypesize{\tiny}
\tablecolumns{11}
\tablecaption{Data for Sample Stars\label{resultstab}}
\tablehead{
&\multicolumn{3}{c}{Position\tablenotemark{a}}&\multicolumn{3}{c}{Velocity (km s$^{-1}$)\tablenotemark{b}}&\colhead{MS Mass\tablenotemark{e}}&\colhead{MS  Lifetime\tablenotemark{e}}&\colhead{T$_{flight}$}&\colhead{Z$_{ej}$}\\
\colhead{Star}&\colhead{R (kpc)}&\colhead{$\theta$ (deg)\tablenotemark{c}}&\colhead{Z  (kpc)\tablenotemark{d}}&\colhead{V$_{\pi}$}&\colhead{V$_{\theta}$}&\colhead{V$_{Z}$}&\colhead{M$_\sun$}&\colhead{Myr}&\colhead{Myr}&\colhead{(km s$^{-1}$)}
}
\startdata
HD 1112&8.2&3.5&-0.3&104$\pm$21&188$\pm$8&6$\pm$3&2.1&800&20$\pm$1&-33$\pm$2\\
HIP 1511&9.8&19.7&-1.7&-263$\pm$42&-167$\pm$33&-301$\pm$104&1.7&1300&7$\pm$13&-311$\pm$100\\
BD -15 115&9.5&26.7&-3.1&49$\pm$19&178$\pm$15&-88$\pm$5&6.8&50&24$\pm$1&-173$\pm$6\\
HD 8323&8.6&3.9&-0.7&-58$\pm$8&192$\pm$7&-10$\pm$6&2.7&400&22$\pm$2&-47$\pm$2\\
BD -7 230&11.8&15.8&-3.9&146$\pm$34&89$\pm$30&37$\pm$8&3.5&250&51$\pm$4&-313$\pm$45\\
Feige 23&9.1&2.1&-0.9&-73$\pm$17&249$\pm$5&-60$\pm$18&2.4&530&13$\pm$3&-79$\pm$14\\
HD 15910&8.8&-2.9&-0.8&5$\pm$2&219$\pm$4&54$\pm$5&2.9&400&61$\pm$4&-69$\pm$4\\
HIP 12320&10.7&4.7&-2.1&53$\pm$14&185$\pm$21&-2$\pm$14&3.7&200&43$\pm$6&-97$\pm$8\\
HD 21305&10.2&-8.9&-2.0&-0$\pm$5&239$\pm$7&-57$\pm$6&6.1&70&24$\pm$1&-98$\pm$4\\
HD 21532&8.7&-3.8&-0.7&40$\pm$5&214$\pm$3&6$\pm$7&3.4&250&28$\pm$3&-49$\pm$2\\
HD 40267&10.3&-13.5&-1.1&168$\pm$20&298$\pm$20&-21$\pm$8&6.3&60&27$\pm$3&-60$\pm$5\\
BD +61 996&14.2&9.5&3.3&36$\pm$24&118$\pm$30&95$\pm$25&8.1&30&29$\pm$5&138$\pm$17\\
HIP 41979&9.6&-3.7&0.9&-118$\pm$20&29$\pm$38&26$\pm$7&6.0&60&22$\pm$3&54$\pm$4\\
HD 233662&12.0&2.8&2.8&6$\pm$7&45$\pm$30&83$\pm$16&6.9&50&27$\pm$3&123$\pm$11\\
HD 237844&10.8&3.9&2.2&8$\pm$10&171$\pm$14&68$\pm$16&5.9&70&25$\pm$4&105$\pm$10\\
BD +16 2114&8.7&-2.6&0.8&61$\pm$10&214$\pm$2&21$\pm$7&3.7&200&19$\pm$2&58$\pm$2\\
BD +38 2182&12.6&-0.4&4.1&130$\pm$31&242$\pm$16&15$\pm$21&6.6&50&53$\pm$8&147$\pm$12\\
Feige 40&8.6&-3.4&1.1&-5$\pm$3&177$\pm$7&58$\pm$5&3.9&200&15$\pm$1&88$\pm$4\\
HD 100340&8.9&-8.0&2.3&16$\pm$9&219$\pm$24&297$\pm$18&10.&20&8$\pm$1&313$\pm$17\\
HD 103376&8.3&-1.1&0.6&-19$\pm$3&256$\pm$5&39$\pm$2&3.0&300&13$\pm$1&59$\pm$2\\
HD 105183&8.7&-6.4&2.9&50$\pm$14&94$\pm$24&-4$\pm$12&4.7&100&42$\pm$3&157$\pm$15\\
HD 106929&8.0&-1.2&0.5&1$\pm$3&211$\pm$3&17$\pm$3&3.1&300&15$\pm$1&45$\pm$1\\
BD +49 2137&9.2&2.8&1.5&119$\pm$20&267$\pm$6&90$\pm$9&3.4&250&14$\pm$1&121$\pm$7\\
BD +36 2268&11.6&1.6&3.8&66$\pm$10&167$\pm$24&29$\pm$4&6.9&50&50$\pm$2&112$\pm$5\\
HD 110166&9.2&1.2&1.5&-43$\pm$5&207$\pm$7&-34$\pm$11&4.2&150&55$\pm$6&80$\pm$5\\
BD +30 2355&8.0&13.7&1.7&-36$\pm$6&93$\pm$19&9$\pm$3&3.2&300&31$\pm$1&99$\pm$2\\
PB 166&9.2&5.9&2.3&292$\pm$56&98$\pm$20&41$\pm$6&5.4&90&24$\pm$2&186$\pm$28\\
Feige 84&5.1&-3.2&3.0&-51$\pm$14&34$\pm$31&137$\pm$9&5.1&90&16$\pm$1&218$\pm$7\\
HD 121968&5.7&-6.6&2.2&86$\pm$19&416$\pm$36&139$\pm$28&7.4&40&13$\pm$2&191$\pm$22\\
HD 123884&5.4&-14.4&2.2&3$\pm$5&151$\pm$16&-9$\pm$10&3.2&250&34$\pm$4&123$\pm$9\\
HD 125924&6.2&-4.5&1.5&-227$\pm$14&125$\pm$10&108$\pm$19&6.9&50&12$\pm$2&141$\pm$16\\
BD +20 3004&5.4&4.8&2.6&115$\pm$26&205$\pm$8&98$\pm$18&3.6&200&17$\pm$2&199$\pm$11\\
HD 137569&6.3&3.8&1.4&-85$\pm$16&169$\pm$9&-102$\pm$16&4.7&100&72$\pm$11&117$\pm$11\\
HD 138503&5.8&-5.9&1.0&12$\pm$4&243$\pm$5&-35$\pm$8&8.1&30&32$\pm$4&81$\pm$5\\
HD 140543&1.8&-50.03&3.1&51$\pm$12&106$\pm$34&19$\pm$17&22.&7&23$\pm$3&260$\pm$44\\
BD +33 2642&8.3&128.12&7.9&293$\pm$76&107$\pm$57&327$\pm$97&10.&20&24$\pm$11&368$\pm$91\\
HD 146813&8.1&10.7&1.5&-5$\pm$7&206$\pm$12&27$\pm$5&8.2&30&25$\pm$1&84$\pm$2\\
HD 149363&6.8&1.6&0.5&-141$\pm$5&189$\pm$16&67$\pm$4&15.&10&8$\pm$1&82$\pm$4\\
BD +13 3224&5.7&12.6&1.6&104$\pm$8&186$\pm$11&72$\pm$17&6.9&50&15$\pm$2&139$\pm$11\\
HD 183899&4.1&13.9&-1.4&-3$\pm$6&244$\pm$7&-5$\pm$13&11.&20&18$\pm$3&-146$\pm$29\\
HD 188618&5.5&12.6&-1.1&-88$\pm$5&287$\pm$8&53$\pm$15&9.7&25&38$\pm$8&-88$\pm$8\\
AC +03 2773&7.77&2.68&-0.2&-17$\pm$4&232$\pm$6&-13$\pm$6&1.6&4300&13$\pm$3&-186$\pm$24\\
HD 206144&5.5&26.7&-2.5&-138$\pm$29&94$\pm$24&-52$\pm$10&9.7&25&25$\pm$2&-135$\pm$11\\
HD 209684&7.0&11.2&-1.2&-65$\pm$3&274$\pm$4&-51$\pm$5&6.8&50&16$\pm$1&-95$\pm$3\\
HD 213781&7.4&7.9&-0.8&9$\pm$2&233$\pm$3&40$\pm$4&4.1&150&38$\pm$2&-76$\pm$3\\
HD 216135&7.2&12.6&-1.3&-200$\pm$30&170$\pm$11&27$\pm$9&4.5&130&46$\pm$6&-64$\pm$7\\
HD 218970&7.2&8.8&-1.1&322$\pm$64&342$\pm$5&-8$\pm$10&5.2&90&19$\pm$2&-106$\pm$7\\
HD 220787&7.7&7.6&-0.7&-5$\pm$2&233$\pm$2&-23$\pm$2&5.7&80&17$\pm$1&-60$\pm$1\\
HD 222040&7.2&10.7&-1.2&23$\pm$11&46$\pm$36&-127$\pm$15&1.1&4600&9$\pm$1&-151$\pm$12\\
\enddata
\tablenotetext{a}{Assumes that the Sun is 8.0 kpc from the galactic center and 0.0 kpc from the galactic plane.}
\tablenotetext{b}{The velocity of the Sun in this frame is:  (V$_{\pi}$, V$_{\theta}$, V$_{Z}$) = (-9 km s$^{-1}$, 232 km s$^{-1}$, 7  km s$^{-1}$)}
\tablenotetext{c}{The angle of separation between the Sun and the star as viewed from the galactic center.  Positive  angles are measured in the direction of galactic rotation.}
\tablenotetext{d}{The distance from the galactic plane to the star in the direction of the north galactic pole.}
\tablenotetext{e}{Assuming the star is on the hydrogen burning main sequence.  Obtained from the tables of \citet{schaller92} using T$_eff$, log(g), and [Fe/H].}
\tablenotetext{f}{AG +03 2773 is probably an F dwarf (see Paper~I).}
\end{deluxetable}

\begin{deluxetable}{lrcclll}
\tablewidth{0pt}
\tabletypesize{\tiny}
\tablecolumns{7}
\tablecaption{Kinematic Scoring \& Final Classifications\label{tab82}\label{tab83}}
\tablehead{
&\colhead{(V$_{\pi}$+V$_{\theta}$)\tablenotemark{a}}&&\colhead{Kinematic}&\colhead{Abundance}\\
\colhead{Star}&\colhead{km s$^{-1}$}&\colhead{(T$_{flight}$/T$_{MS}$)}&\colhead{Score\tablenotemark{b}}&\colhead{Class\tablenotemark{e}}&\colhead{Other\tablenotemark{d}}&\colhead{Verdict\tablenotemark{c}}
}
\startdata 
HD 1112&109&0.03&3&?&&OES?\\
HIP 1511&468&0.05&2&MP (-1)&&OES\\
BD -15 115&65&0.49&2&Disk&&Pop I\\
HD 8323&65&0.06&4&&&Pop I?\\
BD -7 230&196&0.20&2&MP (-.5)&&Pop I?\\
Feige 23&78&0.02&4&&NaD Em; Paper~I, Fig.\ 12&Pop I?\\
HD 15910&5&0.16&4&MP (-.5)&&Pop I\\
HIP 12320&63&0.21&4&Disk?&&Pop I\\
HD 21305&19&0.37&4&OES&&OES\\
HD 21532&41&0.11&4&Disk&&Pop I\\
HD 40267&185&0.43&2&OES&&OES\\
BD +61 996&108&0.86&2&Disk?&&Pop I\\
HIP 41979&225&0.37&2&OES&He EW; Paper~I, Fig.\ 11&OES\\
HD 233662&175&0.58&1&OES?&&OES?\\
HD 237844&50&0.36&4&MP (-.75)&&OES?\\
BD +16 2114&61&0.10&4&&ER Leo&Pop I?\\
BD +38 2182&132&0.99&2&Disk?&&Pop I?\\
Feige 40&44&0.08&4&Disk?&&Pop I\\
HD 100340&16&0.34&4&Disk?&&Pop I?\\
HD 103376&41&0.04&4&Disk?&&Pop I\\
HD 105183&136&0.37&3&OES&&OES\\
HD 106929&9&0.05&4&MP (-.5)&&Pop I\\
BD +49 2137&12&0.06&3&&&Pop I?\\
BD +36 2268&84&1.04&3&OES&&OES\\
HD 110166&45&0.37&4&MP (-.5)&&Pop I?\\
BD +30 2355&132&0.10&3&Disk?&&Pop I\\
PB 166&317&0.28&1&OES&&OES\\
Feige 84&193&0.17&2&Disk?&&Pop I?\\
HD 121968&214&0.32&2&&&U\\
HD 123884&69&0.13&4&MP (-.75)&\citet{bidelman88}&OES?\\
HD 125924&246&0.25&2&&&U\\
BD +20 3004&116&0.08&3&&&Pop I?\\
HD 137569&99&0.64&3&OES?&RV Var; \citet{bolton80}&OES\\
HD 138503&26&1.00&3&Disk&IT Lib; \citet{itlib}&Pop I\\
HD 140543&125&3.08&1&Disk?&&Pop I?\\
BD +33 2642&314&1.10&1&OES?&PN Em; Paper~I, Appendix A&OES\\
HD 146813&15&0.75&3&Disk&&Pop I\\
HD 149363&144&0.65&2&Disk&&Pop I?\\
BD +13 3224&109&0.32&3&&V652 Her; \citet{saio00}&OES\\
HD 183899&24&0.91&3&Disk&&Pop I\\
HD 188618&110&1.58&3&Disk&&Pop I\\
AG +03 2773&40&0.00&4&&&F Dwarf\\
HD 206144&187&1.02&1&Disk&&Pop I?\\
HD 209684&85&0.34&4&Disk?&&Pop I\\
HD 213781&16&0.24&4&Disk?&&Pop I\\
HD 216135&206&0.35&2&Disk?&&Pop I?\\
HD 218970&344&0.22&2&Disk?&&Pop I?\\
HD 220787&14&0.22&4&Disk&&Pop I\\
HD 222040&1752&0.00&2&MP (-1.8)&&OES?\\
\enddata
\tablenotetext{a}{Distance from (V$_{\pi}$,V$_{\theta}$)=(0 km s$^{-1}$, 220 km s$^{-1}$) in the V$_{\pi}$ versus V$_{\theta}$ plane.}
\tablenotetext{b}{See text for explanation of scoring.}
\tablenotetext{c}{``Pop I'' denotes a massive young Population I star ejected from the galactic disk.  ``OES'' denotes an older evolved star.  ``U'' denotes uncertain, there is not enough information to fit the star into the other two categories.  A question mark (?) denotes uncertainty in the classification.}
\tablenotetext{d}{Notes of other factors that were taken into account
  in the classification, i.e. emission features, strong He I features
  , or a variable radial velocity (See Paper~I).} 
\tablenotetext{e}{See Paper~I.  Blank denotes no abundance data. ``Disk'' are abundances consistent with the nearby B stars.  ``MP'' are metal poor abundances that are uniformly depressed without measured abundances for key elements that might show that the star is evolved.  ``OES'' are abundances consistent with patterns expected in older evolved stars.}
\end{deluxetable}

\begin{deluxetable}{llcrrlrrl}
\tablewidth{0pt}
\tabletypesize{\tiny}
\tablecolumns{9}
\tablecaption{Classification of Old Evolved Stars\label{oesclass}}
\tablehead{
&&\colhead{Kinematic}&&&\colhead{Near}&\colhead{vsin(i)}&\colhead{v$_{crit}$\tablenotemark{6}}\\
\colhead{Star}&\colhead{Abundance Comments\tablenotemark{a}}&\colhead{Score}&
\colhead{Log(T$_{eff}$)\tablenotemark{a}}&\colhead{Log(g)\tablenotemark{a}}&
\colhead{HB?}&\colhead{(km s$^{-1}$)}&\colhead{(km s$^{-1}$)}
&\colhead{Class}
}
\startdata 
HD 1112&Al, Si, Fe only, don't agree&3&4.06&4.00&Yes&130&290&Ambiguous?\\
HIP 1511&Fe peak only =-1.3&2&3.95&3.00&Yes&20&170&HB\\
HD 21305&Partial CNO pattern, atomic diffusion&4&4.27&4.00&No&15&290&HB\\
HD 40267&NO up, C down&4&4.23&3.50&No&100&220&Ambiguous\\
HIP 41979&Partial CNO pattern?, atomic diffusion&1&4.25&4.50&Yes&8&380&HB\\
HD 233622&O up?;Z $\sim$ -0.75 &1&4.32&3.50&No&225&220&Ambiguous?\\
HD 237844&Z $\sim$ -0.75&4&4.30&3.75&No&225&250&Ambiguos?\\
HD 105183&N up, C down&3&4.18&3.75&No&70&250&Ambiguous\\
BD +36 2268&N up, C down, Si + Al low&3&4.30&3.75&No&60&250&Ambiguous\\
PB 166&Partial CNO pattern, atomic diffusion&0&4.30&4.75&Yes&5&440&HB\\
HD 123884&Z $\sim$ -0.75 and \citet{bidelman88}&4&4.05&2.5&No&20&120&Post-HB?\\
HD 137569&C down, S (-.41)&3&4.09&2.75&No&20&140&Post-HB\\
BD +33 2642&Al low, S + Ar (-.5), C up, N + O up?&1&4.33&3.25&No&5&190&Post-HB\\
BD +13 3224&Spectra not analyzed&3&4.45&4.00&No&narrow&290&He-Rich Pulsator\tablenotemark{b}\\
HD 222040&Fe peak only =-1.8&2&3.91&3.25&Yes&6&200&HB?\\
\enddata
\tablenotetext{a}{Data from Paper~I.}
\tablenotetext{b}{See \citet{saio00}.}
\tablenotetext{c}{The critical rotational velocity for an evolved star
  of this mass and surface gravity.  See text.}
\end{deluxetable}

\begin{deluxetable}{lrrrr}
\tablewidth{0pt}
\tabletypesize{\small}
\tablecolumns{5}
\tablecaption{Sample Breakdown Compared With Others\label{breakdown}}
\tablehead{
&\colhead{Total}&\colhead{Pop I(\%)}&\colhead{OES(\%)}&\colhead{Unknown}
}
\startdata 
This Study\tablenotemark{a}&48&31(64\%)&15(31\%)&2(4\%)\\
G\&S Stars\tablenotemark{b}&32&23(72\%)&7(25\%)&2(6\%)\\
\\
\citet{magee01}&21&14(67\%)&7(33\%)&\nodata\\
\citet{rolleston97}&25&17(68\%)&7(28\%)&\nodata\\
Ramspeck et al.\tablenotemark{c}&18&14(78\%)&4(22\%)&\nodata\\
\enddata
\tablenotetext{a}{The F dwarf AG +03 2773 has been dropped from the sample in this table.}
\tablenotetext{b}{The stars in our sample taken from \citet{gs74}.}
\tablenotetext{c}{Numbers combined from \citet{ramspeck01a} and \citet{ramspeck01b}.}
\end{deluxetable}

\begin{deluxetable}{lrrrrrrr}
\tablewidth{0pt}
\tabletypesize{\small}
\tablecolumns{8}
\tablecaption{Revised OES Distances And Velocities\label{bhbkin}}
\tablehead{
&&\colhead{Mass}&&\colhead{D}&
\colhead{V$_{\pi}$}&\colhead{V$_{\theta}$}&\colhead{V$_z$}\\
\colhead{Star}&\colhead{Model\tablenotemark{a}}&\colhead{(M$_\sun$)\tablenotemark{a}}&\colhead{M$_V$\tablenotemark{a}}&\colhead{(kpc)}&\colhead{(km s$^{-1}$)}&\colhead{(km s$^{-1}$)}&\colhead{(km s$^{-1}$)}
}
\startdata 
\cutinhead{BHB Stars}
HIP 1511&e64&0.56&0.22&1.73$\pm$0.52&-200$\pm$23&-61$\pm$18&-103$\pm$56\\
HD 21305&z62&0.52&1.52&0.57$\pm$0.17&22$\pm$2&217$\pm$2&-54$\pm$4\\
HIP 41979&z22&0.49&3.05&0.46$\pm$0.14&-42$\pm$5&182$\pm$9&11$\pm$2\\
PB 166&z22&0.49&3.07&0.74$\pm$0.22&94$\pm$17&209$\pm$6&45$\pm$2\\
HD 222040&g14&0.64&0.61&1.09$\pm$0.33&13$\pm$9&77$\pm$31&-117$\pm$13\\
\cutinhead{Post-HB Stars}
HD 123884&e64&0.51&-1.79&1.50$\pm$0.45&-10$\pm$2&194$\pm$6&6$\pm$5\\
HD 137569&j63&0.52&-1.83&0.85$\pm$0.26&-31$\pm$8&200$\pm$3&-67$\pm$7\\
BD +33 2642&\nodata&\nodata&\nodata&3.30$\pm$0.40\tablenotemark{b}&101$\pm$11&90$\pm$11&62$\pm$15\\
BD +13 3224&\nodata&\nodata&\nodata&1.70$\pm$0.02\tablenotemark{c}&60$\pm$2&212$\pm$3&46$\pm$5\\
\cutinhead{Ambiguous}
HD 1112&e64&0.56&+1.51&0.30$\pm$0.09&67$\pm$14&202$\pm$6&6$\pm$3\\
\nodata&\nodata&2.1&+0.63&0.44$\pm$0.13&104$\pm$21&188$\pm$8&6$\pm$3\\
HD 40267&z42&0.51&-0.94&1.34$\pm$0.40&84$\pm$8&243$\pm$8&-16$\pm$3\\
\nodata&\nodata&6.3&-2.71&3.06$\pm$0.92&168$\pm$20&298$\pm$20&-21$\pm$8\\
HD 233662&e64&0.52&-0.21&1.08$\pm$0.32&15$\pm$6&181$\pm$9&46$\pm$8\\
\nodata&\nodata&6.9&-3.06&4.01$\pm$1.20&6$\pm$7&45$\pm$30&83$\pm$16\\
HD 237844&e64&0.51&+0.43&0.85$\pm$0.26&6$\pm$5&216$\pm$4&32$\pm$8\\
\nodata&\nodata&5.9&-2.16&2.94$\pm$0.88&8$\pm$10&171$\pm$14&68$\pm$16\\
HD 105183&z22&0.49&+1.07&0.95$\pm$0.28&9$\pm$4&181$\pm$7&26$\pm$5\\
\nodata&\nodata&4.7&-1.47&3.05$\pm$0.92&50$\pm$14&94$\pm$24&-4$\pm$12\\
BD +36 2268&e64&0.51&+0.38&1.03$\pm$0.31&15$\pm$3&223$\pm$4&34$\pm$2\\
\nodata&\nodata&6.9&-2.49&3.86$\pm$1.16&66$\pm$10&167$\pm$24&29$\pm$4\\
\enddata
\tablenotetext{a}{From tracks of \citet{dorman93} for the following
  model series:  g14 (Y$_HB$ = 0.245; [Fe/H] = -2.26), e64 (Y$_HB$ =
  0.247; [Fe/H] = -1.48), j63 (Y$_HB$ = 0.257; [Fe/H = -0.47), z22
    (Y$_HB$ = 0.288; [Fe/H] = +0.00), z42 (Y$_HB$ = 0.292; [Fe/H] =
    +0.39), z62 (Y$_HB$ = 0.289; [Fe/H] = +0.58).} 
\tablenotetext{b}{Distance taken from \citet{napiwotzki94}.}
\tablenotetext{c}{Distance taken from \citet{jeffery01}.}

\end{deluxetable}

\begin{deluxetable}{c}
\tablewidth{0pt}
\tabletypesize{\small}
\tablecolumns{1}
\tablecaption{Galactic Potential Model Parameters\label{tab66}}
\tablehead{\colhead{\citet{miyamoto75} Disk Potential}}
\startdata 
$\Phi_{disk}=-\frac{GM_{disk}}{\sqrt{R^2+(a+\sqrt{z^2+b^2})^2}}$\\
$M_{disk}=1.0\times 10^{11} \; M_{\sun}$\\
a = 6.5 kpc\\
b = 0.26 kpc\\
\cutinhead{\citet{hernquist90} Bulge/Spheroidal Potential}
$\Phi_{spheroid}=-\frac{GM_{bulge}}{r+c}$\\
$M_{bulge}=3.4\times 10^{10} \; M_{\sun}$\\
c = 0.7 kpc\\
$r = \sqrt{R^2 + z^2}$\\
\cutinhead{Dark Halo Logarithmic Potential}
$\Phi_{halo}=-(v{_{halo}})^2\times ln(r^2+d^2)$\\
$v_{halo}=128 \; km \; s^{-1}$\\
d = 12.0 kpc\\
\enddata
\end{deluxetable}

\begin{deluxetable}{lrrrr}
\tablewidth{0pt}
\tabletypesize{\small}
\tablecolumns{5}
\tablecaption{Comparison with \citet{magee01}\label{tab67}}
\tablehead{
&\multicolumn{2}{c}{Flight Time (Myr)}&
\multicolumn{2}{c}{Ejection Velocity (km s$^{-1}$)}\\
\colhead{Star}&\colhead{Magee}&\colhead{This Work}&\colhead{Magee}&
\colhead{This Work}
}
\startdata 
EC 00321-6320&30&45$\pm$3&58&57$\pm$2\\
EC 00358-1516&32&29$\pm$1&278&226$\pm$18\\
EC 00468-5622&33&39$\pm$2&93&96$\pm$3\\
EC 03342-5243&17&18$\pm$1&145&89$\pm$3\\
EC 05515-6231&21&36$\pm$4&37&39$\pm$4\\
EC 19071-7634&6&6$\pm$1&266&178$\pm$31\\
EC 19337-6734&23&14$\pm$2&48&39$\pm$3\\
EC 19476-4109&52&35$\pm$6&79&81$\pm$9\\
EC 20089-5659&35&21$\pm$3&60&39$\pm$2\\
\enddata
\end{deluxetable}

\begin{deluxetable}{lccccccc}
\tablewidth{0pt}
\tabletypesize{\small}
\tablecolumns{8}
\tablecaption{Alternate Galactic Potential Models\label{tab69}}
\tablehead{
&\colhead{$\Sigma_{disk}$}&\colhead{M$_{disk}$}&\colhead{a\tablenotemark{b}}&
\colhead{M$_{bulge}$}&\colhead{c\tablenotemark{c}}&\colhead{v$_{halo}$}&
\colhead{d\tablenotemark{d}}\\
\colhead{Model}&\colhead{(M$_{\sun}$ pc$^{-2}$)}&
\colhead{(M$_{\sun} \times 10^{11}$)}&\colhead{(kpc)}
&\colhead{(M$_{\sun} \times 10^{10}$)}
&\colhead{(kpc)}&\colhead{(km s$^{-1}$)}&\colhead{(kpc)}
}
\startdata 
Baseline\tablenotemark{a}&95&1.00&6.5&3.4&0.7&128&12.0\\
(A) No Bulge&95&1.00&6.5&0.0&0.0&103&1.0\\
(B) Disk 0.5&47&0.50&6.5&3.4&0.7&113&4.0\\
(C) Disk 0.6&56&0.60&6.5&3.4&0.7&115&5.0\\
(D) Disk 0.7&66&0.70&6.5&3.4&0.7&115&6.0\\
(E) Short Disk&47&0.67&2.6&0.0&0.0&132&6.5\\
\enddata
\tablenotetext{a}{The baseline model is the model in Appendix \ref{galmod}.}
\tablenotetext{b}{Disk scale length}
\tablenotetext{c}{Bulge scale length}
\tablenotetext{d}{Halo scale length}
\end{deluxetable}

\end{document}